\documentclass[a4paper,10pt]{article}
\pdfoutput=1 

\usepackage{jheppub} 

\usepackage[T1]{fontenc} 
\usepackage{lmodern} 

\usepackage{mathabx}
\usepackage{xcolor}
\usepackage{amsmath,amssymb}
\usepackage{mathrsfs,wasysym}
\usepackage{booktabs}
\usepackage{slashed}
\usepackage{cancel}
\usepackage{rotating}
\usepackage{physics}

\newcommand{\as}{\alpha_s}

\newcommand{\xb}{\bar{x}}

\newcommand{\eu}{\mathrm{e}}
\newcommand{\Ab}{\bar{A}}
\newcommand{\Kb}{\bar{K}}
\newcommand{\Cb}{\bar{C}}
\newcommand{\bb}{\bar{b}}
\newcommand{\xfitter}{\texttt{xFitter}}
\newcommand{\MSbar}{\overline{\rm MS}}
\newcommand{\qb}{\bar{q}}

\newcommand{\uda}{\updownarrows}
\newcommand{\dua}{\downuparrows}
\newcommand{\uparr}{\uparrow}
\newcommand{\dwarr}{\downarrow}
\let\originalleft\left
\let\originalright\right
\renewcommand{\left}{\mathopen{}\mathclose\bgroup\originalleft}
\renewcommand{\right}{\aftergroup\egroup\originalright}
\def\beq{\begin{equation}}  
\def\eeq{\end{equation}}
\def\({\left(}
\def\){\right)}
\def\[{\left[}
\def\]{\right]}

\let\oldsubsection\subsection
\renewcommand\subsection[2][\subsectiontoc]{%
  \def\subsectiontoc{#2}%
  \oldsubsection[#1]{\boldmath #2}%
}

\let\oldsubsubsection\subsubsection
\renewcommand\subsubsection[2][\subsubsectiontoc]{%
  \def\subsubsectiontoc{#2}%
  \oldsubsubsection[#1]{\boldmath #2}%
}

\allowdisplaybreaks

\graphicspath{{images/}}

\arxivnumber{2311.08785}

\title{Analysis of HERA data with a PDF parametrization inspired by quantum statistical mechanics}

\author[a]{Marco Bonvini,}
\affiliation[a]{INFN, Sezione di Roma 1, Piazzale Aldo Moro~5, 00185 Roma, Italy}

\author[a]{Franco Buccella,}

\author[b]{Francesco Giuli,}
\affiliation[b]{CERN, 1211 Geneva 23, Switzerland}

\author[c]{Federico Silvetti}
\affiliation[c]{Institute for Particle Physics Phenomenology, Durham University, Durham DH1 3LE, UK}

\preprint{IPPP/23/56}

\emailAdd{marco.bonvini@roma1.infn.it}
\emailAdd{franco.buccella@roma1.infn.it}
\emailAdd{francesco.giuli@cern.ch}
\emailAdd{federico.silvetti@durham.ac.uk}

\abstract{%
We present a determination of the parton distribution functions (PDFs)
of the proton from HERA data using a PDF parametrization 
inspired by a quantum statistical model of the proton dynamics.
This parametrization is characterised by a very small number of parameters,
yet it leads to a reasonably good description of the data,
comparable with other parametrizations on the market.
It may thus provide an
alternative to standard parametrizations,
useful for studying parametrization bias and to possibly simplify the fit procedure
thanks to the small number of parameters.
Interestingly, the model reproduces key physical features,
such as a $\bar d$ distribution larger than $\bar u$,
that HERA data alone are not able to constrain when using more flexible parametrizations.
Moreover, polarized distributions are described in the model by the same parameters
of the unpolarized ones,
giving us the possibility of extracting both types of distributions within the same fit.
}

\begin{document}

\maketitle

\section{Introduction}

Parton distributions functions (PDFs) describe the longitudinal momentum fraction $x$ distributions
of partons (quarks and gluons) inside the proton and they are a key ingredient for
the theoretical description of collisions with protons in the initial state.
For this reason, in the Large Hadron Collider (LHC) era,
a huge effort from both the theory and experimental communities
to improve their determination took place.
PDFs parametrize a low-scale, non-perturbative dynamics of the proton,
and cannot thus be determined using perturbation theory.
Therefore, PDFs are usually fitted from data,
mostly coming from the HERA collider deep inelastic scattering (DIS) experiments,
but also with ever increasing LHC inputs.

PDF fits are performed by several groups~\cite
{H1:2015ubc,Alekhin:2017kpj,Hou:2019efy,Bailey:2020ooq,NNPDF:2021njg,Accardi:2021ysh,ATLAS:2021vod,CMS:2021yzl}
and differ by many aspects:
from the theory description of the data
to the technicalities of the fitting procedure,
from the datasets included in the fit\footnote
{In particular, some PDF fits are based on a global set of experimental data,
  including (collider and fixed-target) DIS, (collider and fixed-target) Drell-Yan,
  jet production, $V$+jet, etc.,
  while other fits are based on more limited sets, e.g.\ HERAPDF~\cite{H1:2015ubc} uses only HERA DIS data.}
to the evaluation of uncertainties.
One distinctive aspect of the various PDF fits is the choice of the functional
form used to parametrize the $x$ dependence of PDFs at the initial scale
(at any other scale PDFs are obtained by perturbative DGLAP evolution).
Some PDF fits use very few parameters, e.g.\ HERAPDF depends on 14 free parameters~\cite{H1:2015ubc},
while other fits use very flexible parametrizations, e.g.\ NNPDF with hundreds of parameters~\cite{NNPDF:2021njg}.
The use of so many parameters in the NNPDF framework is possible thanks
to the application of machine learning techniques to the minimisation of the $\chi^2$.
However, most PDF fitting groups use more traditional techniques,
which are unable to deal with so many parameters.
In these traditional fitting frameworks, having flexible parametrizations
with a small number of parameters is a value.

In this work we consider a parametrization
inspired by a quantum statistical model of the proton dynamics.
This parametrization is characterised by a very small number of parameters.
We use it to fit PDFs from the combined HERA data,
using next-to-next-to-leading order (NNLO) theory without and with the inclusion of small-$x$ resummation,
and it leads to a reasonably good fit despite its limited flexibility,
somewhat comparable with other parametrizations on the market.
We believe that this parametrization can be used as a complement to more standard ones,
both to study parametrization bias and perhaps to facilitate the fit (having few parameters).
For accurate PDF determination, we also consider adding parameters to increase the flexibility
of the parametrization, especially at small $x$. With just two extra parameters,
the quality of the fit becomes competitive with standard parametrizations.

Interestingly, physical features such as a $\bar d$ distribution larger than $\bar u$
come out automatically from the chosen functional form,
even though the HERA data alone are not able to constrain them.
We verify that this property is stable upon inclusion of additional data.
Moreover, the physical model can also describe polarized parton distributions with the same parameters.
If we trust the model, this feature gives the possibility to fit simultaneously polarized and unpolarized distributions
without the need of introducing new degrees of freedom.
We investigate this possibility finding very promising results.

The paper is structured as follows.
In section~\ref{sec:model} we introduce the parametrization from the statistical model.
In section~\ref{sec:setup} we discuss the setup of our fit and introduce a benchmark fit with parametrization \`a la HERAPDF.
In section~\ref{sec:fit} we test the parametrization against HERA data and compare with our benchmark and with other public PDFs.
In section~\ref{sec:flex} we introduce a more flexible version of the parametrization which improves the fit quality
and study its model uncertainty and its comparison with other parametrizations.
In section~\ref{sec:implications} we discuss our result in view of the physical model behind it, and consider possible implications and future directions.
We conclude in section~\ref{sec:conclusions}.

\section{The PDF parametrization from the statistical model}
\label{sec:model}

A field-theoretical computation of PDFs requires dealing with non-perturbative dynamics,
and it is therefore very difficult to achieve.
Even numerical techniques like lattice QCD are not (yet\footnote
{Attempts to determine PDFs in lattice QCD make use of alternative definitions of PDFs,
  denoted quasi-PDFs~\cite{Ji:2013dva} and pseudo-PDFs~\cite{Radyushkin:2017cyf},
  which are well defined in Euclidean space-time and are related in some limit to ordinary light-cone PDFs.})
able to satisfactorily determine PDFs, essentially because PDFs are defined in terms of bilocal operators
separated by a light cone distance which cannot be described on a Euclidean lattice~\cite{Lin:2017snn}.

Usually, PDFs are determined by fitting them to data using an arbitrary parametrization
of the $x$ dependence of the PDFs at an ``initial'' scale $\mu_0$, typically chosen
at the border between perturbative and non-perturbative QCD ($\mu_0\sim1$~GeV).
PDFs are then evolved at different scales solving the perturbative DGLAP equation~\cite{Altarelli:1977zs,Gribov:1972ri,Dokshitzer:1977sg}.
The goal in choosing the parametrization is usually to minimise the bias with sufficiently flexible functional forms
while at the same time keeping the number of parameters small enough
(for better numerical performances and to avoid overfitting).
A notable exception is the parametrization used by the NNPDF collaboration,
which uses neural networks to parametrize PDFs with hundreds of parameters:
this makes any bias completely negligible, but requires the use of sophisticated
machine learning techniques to perform the fit and to determine the uncertainties~\cite{NNPDF:2017mvq,NNPDF:2021njg}.

In Ref.~\cite{Bourrely:1993wq} (see also Refs.~\cite{Buccella:1996kb,Bourrely:2001du,Bourrely:2002xm,Bourrely:2005kw,Bourrely:2005tp,Bourrely:2010ng,Bourrely:2013yti,Bourrely:2013qfa,Bourrely:2014uha,Buccella:2014wpa,Bourrely:2015kla,Bourrely:2018yck,Soffer:2019gbb,Buccella:2019yij,Bourrely:2020izp,Bourrely:2022mjf,Buccella:2022tmb,Bellantuono:2022hqp,Bourrely:2023yzi,Silvetti:2024fdi})
a different approach has been proposed,
where the functional form of the PDFs are obtained through a statistical model of
the parton dynamics in the proton.
The model assumes that the partons could be treated as massless particles
forming an ideal quantum gas at equilibrium at the initial scale $\mu_0$ in a finite volume,
characterised by an effective temperature.
Working in the infinite momentum frame (\`a la Feynman), the transverse degrees of freedom
can be neglected and the dynamics can be described in terms of the longitudinal momentum fraction $x$.
As a result, the model predicts a (very biased) functional form for the PDFs
in terms of very few parameters.

Because the model does not take into account the QCD interaction,
information like the factorization scheme is absent in the parametrization.
This makes the model clearly incomplete.
For instance, the resulting PDFs do not contain information on the scale at which they are supposed to be computed,
as the scale dependence is a consequence of the factorization of collinear singularities in QCD.
We thus do not expect the model to give a reliable description per se,
but we want to investigate if it can provide a suitable baseline for
a parametrization.

To this end, the plain model must be supplemented with ``phenomenological'' modifications.
Some of them were introduced already in the original publications~\cite{Bourrely:1993wq,Buccella:1996kb,Bourrely:2001du},
others have been considered later on~\cite{Bourrely:2005kw,Bourrely:2005tp,Bourrely:2010ng,Bourrely:2013yti,Bourrely:2013qfa,Bourrely:2014uha,Buccella:2014wpa,Bourrely:2015kla,Soffer:2019gbb,Buccella:2019yij,Bellantuono:2022hqp}.
Crucially, the various incarnations of the model describe separately the individual
polarizations of quarks, thus providing in principle a description of polarized and unpolarized PDFs
in terms of the same parameters.

In this work, we will consider the simple parametrization proposed in Ref.~\cite{Bourrely:2001du}.
More flexible functional forms depending on more parameters may provide a better description of the data,
but in this analysis we want to keep the number of parameters as small as possible.
Let us first introduce the function
\beq\label{eq:hpm}
h_\pm(x;b,X) = \frac{x^b}{\exp(\frac{x-X}{\xb})\pm1}, 
\eeq
which descends from Fermi-Dirac ($h_+$) and Bose-Einstein ($h_-$) distributions,
supplemented by a phenomenological power term $x^b$ to describe the small-$x$ asymptotic behaviour.
Within the model, all PDFs can be written as linear combination of this function
with different values of the parameters $b$ and $X$,
the latter representing a ``chemical potential''.
The parameter $\xb$ plays the role of a ``temperature'',
and it is common to all PDFs.
For this reason, we do not write it explicitly as an argument.

Let us start with the quark PDFs.
Being spin-$\frac12$ particles, quarks follow a Fermi-Dirac distribution,
and are thus described in terms of the $h_+$ function.
The PDFs at the initial scale $\mu_0$
for each polarization (indicated with an up or down arrow)
are given by
\begin{subequations}
  \label{eq:param-q}
  \begin{align}
    x q^\uda (x, \mu_0^2) &= A C_q^\uda h_+(x; b, X_q^\uda) + \tilde{A} h_+(x; \tilde b,0)  , \\
    x \qb^\uda (x, \mu_0^2) &= \Ab \Cb_q^\uda h_+(x; \bb, -X_q^\dua) + \tilde{A} h_+(x; \tilde b,0),
  \end{align}
\end{subequations}
where $C_q^\uda$, $\bar C_q^\uda$ and $X_q^\uda$
are parameters depending on the quark species
and polarization, while $b$, $\bb$, $\tilde b$ as well as the normalizations $A$, $\bar A$ and $\tilde A$
are flavour-independent parameters.\footnote
{Note that as far as the parameters $C_q^\uda$ and $\bar C_q^\uda$
    are unconstrained, the normalizations $A$ and $\bar A$ are redundant as they can be reabsorbed
    in a redefinition of those parameters.
    However, we will see that simple incarnations of the model constrain the values of $C_q^\uda$ and $\bar C_q^\uda$
    so that the normalizations $A$ and $\bar A$ are no longer redundant.}
In each equation the first term is of ``valence nature'', and dominates at high $x$.
The second term, called ``diffractive'' term in the original literature,
is identical for all quark flavours and helicities and represents a contribution of ``sea nature'',
which thus is expected to dominate at small $x$.
The unpolarized quark distribution is just the sum of the distributions with opposite helicities.
For example, for an antiquark we have
\begin{subequations}\label{eq:PDFpar}
  \begin{align}\label{eq:PDFparqbar}
    x \qb (x, \mu_0^2) &\equiv x\qb^\uparr (x, \mu_0^2) + x\qb^\dwarr (x, \mu_0^2) \nonumber \\
              & = \Ab \[\Cb_q^\uparr h_+(x;\bb,-X_q^{\dwarr}) + \Cb_q^\dwarr h_+(x;\bb,-X_q^\uparr) \]
                + 2 \tilde{A} h_+(x;\tilde b,0) ,
\end{align}
while for a valence quark distribution $q_v=q-\bar q$ we have
\begin{align}\label{eq:PDFparqv}
  x q_v (x, \mu_0^2) &\equiv xq^\uparr (x, \mu_0^2) + xq^\dwarr (x, \mu_0^2) - x\qb^\uparr (x, \mu_0^2) - x\qb^\dwarr (x, \mu_0^2) \\
            & =  A\[C_q^\uparr h_+(x; b, X_q^\uparr) + C_q^{\dwarr} h_+(x; b, X_q^{\dwarr})\]
            - \Ab \[\Cb_q^\uparr h_+(x;\bb, -X_q^{\dwarr}) + \Cb_q^\dwarr h_+(x; \bb, -X_q^\uparr)\]. \nonumber
\end{align}
Note that the valence distribution does not depend anymore on the diffractive (sea) term.
Finally, the gluon follows a Bose-Einstein distribution with vanishing potential (i.e.\ a Planck distribution)
\begin{equation}\label{eq:gluonPDF}
  x g(x, \mu_0^2) = A_g h_-(x;b_g,0)
\end{equation}
\end{subequations}
depending on a new normalization $A_g$ and an exponent $b_g$.\footnote
{Note that we do not consider separate polarizations for the gluon,
  as suggested in the original literature for the model~\cite{Bourrely:2001du,Bourrely:2002xm}.
  This is perhaps not ideal, as there is experimental evidence that gluons
  carry a non-zero polarization contributing to the proton spin~\cite{deFlorian:2014yva}.
  Phenomenological extensions of the model accounting for the gluon polarization have been discussed
  in the literature~\cite{Bourrely:2014uha,Bourrely:2015kla}, however it is not clear
  whether these extension are in agreement with the model assumptions.
  We will discuss gluon polarization with greater detail in section~\ref{sec:pol}.
  When we consider unpolarized PDFs only, our assumption does not represent a severe limitation.}

We immediately observe that the gluon distribution is very limited,
as it depends only on 3 parameters ($\xb,b_g,A_g$).
Differently, quark distributions depend on many more parameters.
However, there are some relations that constrain some of them, as we shall now see.

Let us start from the asymptotic behaviours.
At small $x$, we expect the gluon and sea quark to behave (grow) in the same way.
The gluon grows as $xg(x)\sim x^{b_g-1}$,
while the quarks are dominated at small $x$ by the sea term\footnote
{The ``valence'' terms depending on $b$ and $\bar b$ contribute to the valence PDFs which cannot grow at small $x$,
so we expect $b$ and $\bb$ to be positive.}
and thus behave as $xq(x)\sim x^{\tilde b}$.
Therefore, we find the relation
\beq
b_g = \tilde b+1
\eeq
that allows us to remove one of the two parameters.

We now focus on the parameters $C_q^\uda, \bar C_q^\uda$.
In the statistical model, they are not independent parameters,
rather they are related to one another and possibly with other parameters of the model.
In Refs.~\cite{Bourrely:2005tp,Bourrely:2013yti,Buccella:2014wpa,Buccella:2019yij}
the original model has been extended to describe transverse degrees of freedom.
After integrating over transverse momentum to get collinear PDFs,
the coefficients $C_q^\uda, \bar C_q^\uda$ are given by
\begin{equation}\label{eq:CudY}
C_q^\uda = \log(1 +\eu^{Y_q^\uda}) \qquad \, \qquad \Cb_q^\uda = \log(1 +\eu^{-Y_q^\dua}) 
\end{equation}
in terms of the parameters $Y_q^\uda$, denoted ``transverse potentials''.
Note that there are two independent $Y$'s for each quark flavour, fixing the four $C$'s for each quark.
We observe that in previous studies, e.g.\ Ref.~\cite{Bourrely:2001du},
the $C$ parameters were simply given in terms of the chemical potentials
$X_q^\uda$ through the relations
\begin{equation}\label{eq:CudX}
C_q^\uda = X_ q^\uda, \qquad \, \qquad \Cb_q^\uda = \frac1{X_ q^\dua}
\end{equation}
(note that in the second equation the order of the arrows changes).
In this way, these four degrees of freedom for each quark flavour are completely fixed,
leaving a parametrization that is rather constrained.
The choice Eq.~\eqref{eq:CudX} of Ref.~\cite{Bourrely:2001du} was justified
by the agreement with data,
and later observed~\cite{Buccella:2014wpa,Buccella:2019yij} to be in decent agreement
with the parametrization Eq.~\eqref{eq:CudY}, up to a rescaling of $A$ and $\bar A$.
In this work, we only consider the simplest model, i.e.\ we adopt Eq.~\eqref{eq:CudX}
in order to reduce the number of parameters to a minimum,
keeping in mind that more flexibility can be achieved by adopting Eq.~\eqref{eq:CudY} instead.

Finally, we have to take into account the sum rules.
Specifically we have two quark number sum rules and the momentum sum rule,
that allow us to fix three additional parameters.
We choose them to be the three normalizations $A,\bar A, A_g$.
We stress that, differently from standard parametrizations,
$A$ and $\bar A$ are not directly the normalizations of $u$ and $d$ valence distributions.
This makes the implementation of the sum rules in the fitting code not straightforward.
We give technical details in Appendix~\ref{sec:sumrules}.

One peculiar feature of the PDF parametrization Eqs.~\eqref{eq:PDFpar}
is that it does not vanish at $x=1$, as all other PDF parametrizations
on the market do (to our knowledge).
In particular the function $h_\pm$ in Eq.~\eqref{eq:hpm} in $x=1$ becomes
\beq\label{eq:QSx=1}
h_\pm(1;b,X) = \frac1{\exp\(\frac{1-X}{\xb}\)\pm1}.
\eeq
As we shall see, in the fits $X<1$ always, thus this function is exponentially suppressed.
The suppression is rather strong, thanks to the value of $\bar x\sim 0.1$
which is common to all fits with this parametrization,
and to the fact that the largest value of $X$ from the fit is smaller than 0.5.
So practically the resulting PDFs are indistinguishable from zero in $x=1$.
Moreover, we recall that $x=1$ corresponds to the elastic scattering limit,
which is no longer described by the QCD factorization theorem.

We conclude the section by counting the number of free parameters that we have in our fit.
As we will only perform a fit to HERA data, we do not parametrize the strange distribution independently,
as the data are not sufficiently powerful to distinguish it from the $\bar d$ distribution.
We thus take it to be a fixed fraction of $\bar d$ distribution,
\begin{equation}
  s(x, \mu_0^2)=\bar s(x, \mu_0^2) = \frac{f_s}{1-f_s} \bar{d}(x, \mu_0^2) ,\qquad f_s=0.4, 
\end{equation}
which is a standard choice adopted by HERAPDF~\cite{H1:2015ubc}.
We are thus left with 5 PDFs to fit, i.e.\ $u_v, d_v, \bar u, \bar d, g$.
According to the parametrizations given above, the free parameters to be fitted are
\beq
\xb, b, \bb, \tilde b, \tilde A, X^\uparr_u, X^\dwarr_u, X^\uparr_d, X^\dwarr_d,
\eeq
for a total of 9 parameters.
For comparison, the default HERAPDF parametrization has 14 free parameters.

\section{Setup of the fit and benchmark}
\label{sec:setup}

Having established the form of the parametrization that we want to use, we now discuss
the setup of our fit.
We use the public \xfitter\ toolkit, using a setup that is close to the one used
for the determination of HERAPDF2.0~\cite{H1:2015ubc}, with some notable differences:
\begin{itemize}
\item 
  First of all, the paper~\cite{Bourrely:2001du} where we take our PDF parametrization
  advocates that it should be used at the initial scale $\mu_0=2$~GeV.
  We therefore consider this scale as our default parametrization scale,
  which is higher than the HERAPDF2.0 scale which is $1.38$~GeV.

\item
  In order to avoid backward evolution, we then keep data only above 2~GeV,
  namely we have to cut out the $Q^2=3.5$~GeV$^2$ bin of the HERA dataset.

\item
  For the same reason, since we want to generate the charm PDF perturbatively,
  we have to raise the charm matching scale $\mu_c$ above the parametrization scale~\cite{xFitterDevelopersTeam:2017fzy}.
  In our work we set it to $\mu_c=1.38m_c=2.01$~GeV, with $m_c=1.46$~GeV.

\item
  To implement the displaced charm threshold, we have to use the \texttt{APFEL} evolution code~\cite{Bertone:2013vaa}
  rather than the default \texttt{QCDNUM} code~\cite{Botje:2010ay}, as only the former implements this feature.
  This implies that we have to change the variable flavour number scheme from TR~\cite{Thorne:1997ga,Thorne:2006qt,Thorne:2012az}
  to FONLL~\cite{Forte:2010ta}.
  In practice, in our
  NNLO fits we use FONLL-C~\cite{Forte:2010ta}.
\end{itemize}
The other settings of fit (masses, couplings, $\chi^2$ definition, minimization strategy, etc) are
kept as in HERAPDF2.0~\cite{H1:2015ubc}.

We now present a HERAPDF-like NNLO fit with these settings, using the default HERAPDF2.0 parametrization,
and compare it with the public HERAPDF2.0.
The default HERAPDF2.0 parametrization is given by~\cite{H1:2015ubc}
\begin{subequations}\label{eq:HERApar}
\begin{align}
  xg(x,\mu_0^2) &= A_g\,x^{B_g} (1-x)^{C_g} -A_g'\,x^{B_g'} (1-x)^{25}\\
  xu_v(x,\mu_0^2) &= A_{u_v}\,x^{B_{u_v}} (1-x)^{C_{u_v}} \Big[1+E_{u_v}x^2\Big] \\
  xd_v(x,\mu_0^2) &= A_{d_v}\,x^{B_{d_v}} (1-x)^{C_{d_v}} \\
  x\bar u(x,\mu_0^2) &= A_{\bar u}\,x^{B_{\bar u}} (1-x)^{C_{\bar u}} \Big[1+D_{\bar u}x\Big] \\
  x\bar d(x,\mu_0^2) &= A_{\bar u}\,x^{B_{\bar u}} (1-x)^{C_{\bar d}}.
\end{align}
\end{subequations}
Our HERAPDF-like fit will serve as a baseline for our next studies with the parametrization presented in Sect.~\ref{sec:model}.
We choose the HERAPDF parametrization for this baseline because it is the simplest among the mainstream PDF parametrizations,
using the smallest set of parameters (14 free parameters in total).
Any other mainstream parametrization on the market has more parameters and it is therefore expected to be more flexible
and possibly lead to higher fit quality.

\begin{table}
\centering
\begin{tabular}{lcc}
  Contribution to $\chi^2$ &HERAPDF2.0 &  Our HERAPDF-like fit  \\
  \midrule
  subset NC $e^+$ 920     & $444/377$   & $415/363$   \\
  subset NC $e^+$ 820     & $ 66/ 70$   & $ 66/ 68$   \\
  subset NC $e^+$ 575     & $219/254$   & $217/249$   \\
  subset NC $e^+$ 460     & $217/204$   & $213/200$   \\
  subset NC $e^-$         & $219/159$   & $214/159$   \\
  subset CC $e^+$         & $ 45/ 39$   & $ 45/ 39$   \\
  subset CC $e^-$         & $ 56/ 42$   & $ 56/ 42$   \\
  correlation term + log term     & $91 + 5$ & $91 + 15$ \\
  \bf\boldmath Total $\chi^2/\rm{d.o.f.}$  &\boldmath $1363/1131$ &\boldmath $1333/1106$ \\
\end{tabular}
\caption{Total $\chi^2$ per degrees of freedom (d.o.f.)
  and the partial $\chi^2$ per number of data points (n.d.p.)
  of each subset of the inclusive HERA dataset,
  for HERAPDF2.0 and a HERAPDF-like fit obtained with the new setting introduced here.}
\label{tab:chi2newsetting}
\end{table}

We start by showing in Table~\ref{tab:chi2newsetting} the $\chi^2$
breakdown for the two fits.
For each subset the contribution to the $\chi^2$ over the number of
data points is shown, as well as the contributions to the $\chi^2$
from the correlations and the logarithmic term (see Ref.~\cite{H1:2015ubc} for
the definition and meaning of these pieces).
Our fit has a total $\chi^2$ which is smaller by 30 units, which is
compatible (within statistical fluctuations) with the reduction of
datapoints by 25 units.
The small improvement is likely due to the better description of the
$E=920$~GeV dataset which contains the small-$x$ data,
which are not well described by fixed-order perturbation
theory~\cite{H1:2015ubc,Harland-Lang:2016yfn,Abt:2016vjh,Ball:2017otu,xFitterDevelopersTeam:2018hym}.
The cut bin at $Q^2=3.5$~GeV$^2$ contains the data at smallest $x$,
thus its absence in the new fit leads to a 29 units smaller $\chi^2$
of the $E=920$~GeV dataset with just 14 less datapoints.

\begin{figure}[t]
  \centering
  \includegraphics[width=0.328\textwidth,page=1]{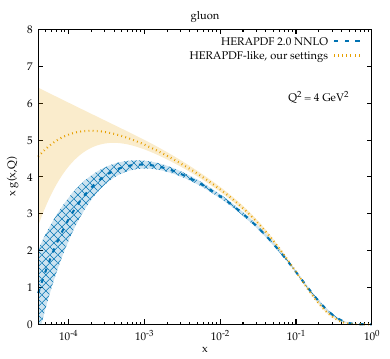}
  \includegraphics[width=0.328\textwidth,page=3]{PDF_comparison__HERAPDF.pdf}
  \includegraphics[width=0.328\textwidth,page=5]{PDF_comparison__HERAPDF.pdf}\\
  \includegraphics[width=0.328\textwidth,page=2]{PDF_comparison__HERAPDF.pdf}
  \includegraphics[width=0.328\textwidth,page=4]{PDF_comparison__HERAPDF.pdf}
  \includegraphics[width=0.328\textwidth,page=6]{PDF_comparison__HERAPDF.pdf}
  \caption{Comparison of the original HERAPDF2.0 fit (dashed blue) with the one with our modified settings (dotted yellow)
    for the gluon, total singlet, $\bar u$, $\bar d+\bar s$, $u_v$ and $d_v$ PDFs.
    The uncertainty shown is only the ``experimental'' one, namely the one coming from the uncertainty on the parameters determined from the fit.}
  \label{fig:PDFnewsetting}
\end{figure}

We now show the effect of the new settings to the PDFs.
In Figure~\ref{fig:PDFnewsetting} we show a comparison of HERAPDF2.0
and our new fit with HERAPDF parametrization
at the scale $Q=2$~GeV for the gluon, total quark singlet, $\bar u$,
$\bar d+\bar s$, $u_v$ and $d_v$ PDFs.
We observe that the valence distributions as well as the medium-large
$x$ behaviour of all PDFs remain almost unchanged in the two fits.
Differences are instead present in the small-$x$ region,
for the sea contribution to the quarks and more markedly to the gluon.
These differences are certainly due, at least in part, by the smaller
dataset,
and in particular by the absence of the small-$x$ data of the
$Q^2=3.5$~GeV$^2$ bin, which also leads to an increased uncertainty in
the gluon PDF at small $x$.\footnote
{On top of this, there is a technical difference in the way uncertainties are calculated.
  HERAPDF2.0 uses the Pumplin procedure~\cite{Pumplin:2000vx},
  while we use the simpler HESSE approach of MINUIT~\cite{James:1975dr}
  which may lead to larger uncertainties.}

\section{Fit with the new parametrization}
\label{sec:fit}

We now consider the parametrization described in
section~\ref{sec:model}.
We perform a fit with the same settings described above, simply
changing the PDF parametrization, that we dub QSPDF.
Comparing it with the HERAPDF-like fit just discussed in
section~\ref{sec:setup} we are able to test the effect of the
different parametrization alone, disentangled from any other effect.

\begin{table}
\centering
\begin{tabular}{lccc}
  & QSPDF &  QSPDF & HERAPDF-like  \\
  Contribution to $\chi^2$ & (NNLO) &  (NNLO+NLL$x$) & (NNLO+NLL$x$)  \\
  \midrule
  subset NC $e^+$ 920      & $452/363$   & $447/363$   & $408/363$ \\
  subset NC $e^+$ 820      & $ 71/ 68$   & $ 66/ 68$   & $ 63/ 68$ \\
  subset NC $e^+$ 575      & $224/249$   & $229/249$   & $216/249$ \\
  subset NC $e^+$ 460      & $221/200$   & $231/200$   & $218/200$ \\
  subset NC $e^-$          & $222/159$   & $225/159$   & $219/159$ \\
  subset CC $e^+$          & $ 46/ 39$   & $ 48/ 39$   & $ 46/ 39$ \\
  subset CC $e^-$          & $ 61/ 42$   & $ 61/ 42$   & $ 54/ 42$ \\
  correlation term + log term   & $98-11$      & $88-28$     & $80+1$ \\
  \bf\boldmath Total $\chi^2/\rm{d.o.f.}$  &\boldmath $1384/1111$ &\boldmath $1369/1111$ &\boldmath $1304/1106$ \\
\end{tabular}
\caption{Same as table~\ref{tab:chi2newsetting}, showing the $\chi^2$ breakdown for QSPDF at NNLO and NNLO+NLL$x$ as well as our HERAPDF-like fit at NNLO+NLL$x$.}
\label{tab:chi2QSPDF}
\end{table}

The $\chi^2$ breakdown for this new fit is shown in
Table~\ref{tab:chi2QSPDF},
to be compared with the last column of Table~\ref{tab:chi2newsetting}.
We immediately observe an overall deterioration of the fit quality,
with the $\chi^2$ increasing by 51 units, from 1333 to 1384.
The number of degrees of freedom also increases slightly (5 units) due
to the smaller number of parameters in the QSPDF parametrization, but
this is not enough to explain the increase in the $\chi^2$.

Looking carefully at the tables, we see that the dataset exhibiting the
largest deterioration is again the $E=920$~GeV dataset, namely the one
containing the majority of small-$x$ data.
We suspect indeed that the origin of the large $\chi^2$ comes from a
bad description of the low-$x$ data, which is in turn due to the
limited flexibility of the QSPDF parametrization at small $x$, in
particular for the gluon.
We will come back to this point later.

\begin{figure}[t]
  \centering
  \includegraphics[width=0.328\textwidth,page=1]{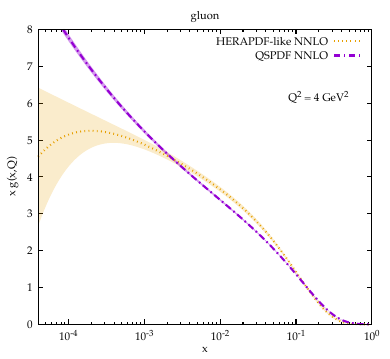}
  \includegraphics[width=0.328\textwidth,page=3]{PDF_comparison__QSPDF_HERAPDF.pdf}
  \includegraphics[width=0.328\textwidth,page=5]{PDF_comparison__QSPDF_HERAPDF.pdf}\\
  \includegraphics[width=0.328\textwidth,page=2]{PDF_comparison__QSPDF_HERAPDF.pdf}
  \includegraphics[width=0.328\textwidth,page=4]{PDF_comparison__QSPDF_HERAPDF.pdf}
  \includegraphics[width=0.328\textwidth,page=6]{PDF_comparison__QSPDF_HERAPDF.pdf}
  \caption{Comparison of QSPDF (dot-dashed purple) with our HERAPDF-like fit (dotted orange)
    for the gluon, total singlet, $\bar u$, $\bar d+\bar s$, $u_v$ and $d_v$ PDFs, showing experimental uncertainty band.}
  \label{fig:QSPDF}
\end{figure}

In Figure~\ref{fig:QSPDF} we show the comparison of our HERAPDF-like
fit with the QSPDFs.
There are some marked differences between the two PDF sets.
Starting with the quarks, we observe a small distortion of the $u_v$
distribution below the peak, with QSPDFs being smaller at medium $x$
and larger at small $x$.
A similar but bigger effect is present on the $d_v$ distribution as well,
where also the height of the peak is smaller in the QSPDF fit.
The anti-quark PDFs behave the same at small $x$,
while at medium-large $x$ the QSPDFs for the total singlet and the
$\bar d+\bar s$ distribution are slightly larger compared to the PDF
uncertainty.
Finally, a big difference is present in the gluon PDF from medium to
low $x$, with the QSPDF gluon rising at small $x$ compared to the
HERAPDF-like gluon which bends down at $x\sim10^{-4}$.
Moreover, the PDF uncertainty of the QSPDFs is very small everywhere,
especially in the gluon where instead the HERAPDF parametrization
gives a much larger uncertainty in the small-$x$ region.

These differences, especially in the gluon, are due to the very
constrained parametrization which limits its flexibility.
However, before drawing conclusions,
it is instructive to compare the QSPDFs with other PDFs on the market.
We consider the NNPDF3.0 set that has been obtained fitting only HERA
data~\cite{Ball:2014uwa},
which is a dataset very similar to what we are using here,
and the NNPDF4.0 set that has been obtained fitting only DIS
data~\cite{NNPDF:2021njg},
which contains more data but it is still closer to our dataset than a
global fit.

\begin{figure}[t]
  \centering
  \includegraphics[width=0.328\textwidth,page=1]{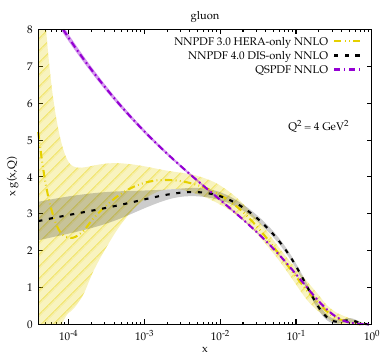}
  \includegraphics[width=0.328\textwidth,page=3]{PDF_comparison__QSPDF_others.pdf}
  \includegraphics[width=0.328\textwidth,page=5]{PDF_comparison__QSPDF_others.pdf}\\
  \includegraphics[width=0.328\textwidth,page=2]{PDF_comparison__QSPDF_others.pdf}
  \includegraphics[width=0.328\textwidth,page=4]{PDF_comparison__QSPDF_others.pdf}
  \includegraphics[width=0.328\textwidth,page=6]{PDF_comparison__QSPDF_others.pdf}
  \caption{Comparison of QSPDF (dot-dashed purple) with NNPDF3.0 HERA-only (dot-dot-dashed yellow) and NNPDF4.0 DIS-only (dashed black)
    for the gluon, total singlet, $\bar u$, $\bar d+\bar s$, $u_v$ and $d_v$ PDFs.
    The NNPDF uncertainty band covers various sources of uncertainty, including those coming from parametrization choice.}
  \label{fig:QSPDFothers}
\end{figure}

The plots are shown in Figure~\ref{fig:QSPDFothers}.
We immediately notice that the NNPDF uncertainties are larger than QSPDFs,
generally due to the very flexible parametrization, with the ones of
the older NNPDF3.0 fit being larger than the newer NNPDF4.0 fit,
partly due to the larger dataset and partly to the improved fitting
methodology in the latter.
Within uncertainties, there is a sufficiently good agreement between
QSPDFs and the NNPDF sets for the total singlet, the $\bar d+\bar s$
and the valence distributions. In particular, we notice that the $u_v$
and $d_v$ PDFs of the QSPDF set have a shape very similar to the NNPDF
ones, despite the differences with the HERAPDF-like fit.
Because of the unbiased nature of the NNPDF parametrization, we thus
conclude that the $u_v$ and $d_v$ distributions are probably better
described by the QSPDF parametrization than by the HERAPDF parametrization.\footnote
{To confirm this, we have tried to fit directly the $u_v$ and $d_v$ distributions from the QSPDF set
  using the HERAPDF parametrization Eq.~\eqref{eq:HERApar}, noticing that indeed the HERAPDF parametrization
  is not able to reproduce the shape of the QSPDF valence distributions in the medium/small-$x$ region.
  Conversely, the QSPDF parametrization is able to reproduce the valence distributions of our HERAPDF-like fit,
  except for the high-$x$ tail which is however mostly unconstrained by data.
  Note that adding more polynomial contributions in the HERAPDF parametrization of the valence PDFs,
  as done in many PDF studies (see e.g.~\cite{ATLAS:2021vod}), will likely give enough flexibility to reproduce a QSPDF valence shape.}

In contrast, the $\bar u$ distribution in QSPDF is somewhat higher at
medium-small $x$ than both NNPDF predictions.
This behaviour is dictated by the sea term of the QSPDF parametrization,
which is in turn linked to the gluon.
The gluon is very different from the NNPDF one, as the latter
tends to flatten below $x\sim10^{-2}$ in both fit versions, while the
QSPDF gluon keeps growing at small $x$.
This behaviour of the QSPDF gluon is a consequence of its very
constrained functional form, which is not able to reproduce a shape
similar to the NNPDF one (and not even of the HERAPDF-like fit).

We have thus found various hints that the gluon parametrization Eq.~\eqref{eq:gluonPDF}
is not sufficient to accurately describe the data at small $x$.
For this reason, we will consider in section~\ref{sec:flex} a more flexible parametrization for the gluon PDF.
However, we must recall that the shape of the gluon PDF at small $x$ is strongly dependent on the perturbative order of the fit,
due to the presence of enhanced logarithmic terms in the perturbative ingredients that make their perturbative expansion
unstable at small $x$.
This instability can be cured by resumming these logarithms to all orders~\cite
{Salam:1998tj,Ciafaloni:1999yw,Ciafaloni:2003kd,Ciafaloni:2003rd,Ciafaloni:2007gf,Ball:1995vc,Ball:1997vf,Altarelli:2001ji,Altarelli:2003hk,Altarelli:2005ni,Altarelli:2008aj,Thorne:1999sg,Thorne:1999rb,Thorne:2001nr,White:2006yh,Rothstein:2016bsq,Catani:1990xk,Catani:1990eg,Catani:1994sq},
leading to predictions that are more reliable in that region.
Interestingly, adding the resummation of small-$x$ logarithms in PDF fits
leads to a gluon PDF that rises at small $x$ at small scales~\cite{Ball:2017otu,xFitterDevelopersTeam:2018hym,Bonvini:2019wxf}.
Therefore, it may be possible that the QSPDF is able to give a better description of
the gluon PDF when small-$x$ resummation is turned on.

To verify this, we have performed fits to the HERA data with small-$x$ resummation at
next-to-leading logarithmic (NLL$x$) accuracy
(included in \xfitter\ through \texttt{APFEL} interfaced to the \texttt{HELL}
resummation code~\cite{Bonvini:2016wki,Bonvini:2017ogt,Bonvini:2018xvt,Bonvini:2018iwt}, version 3.0)
using both the QSPDF and the HERAPDF parametrizations.
The fit quality is reported in the last two columns of table~\ref{tab:chi2QSPDF}.
For QSPDFs we observe a reduction of the $\chi^2$ from NNLO to NNLO+NLL$x$ of 15 units,
which is significant but not substantial.
Conversely, in our HERAPDF-like fit the $\chi^2$ reduces by 29 units, which is more notable
given that it was already lower than the QSPDF one.
These numbers confirm~\cite{Ball:2017otu,xFitterDevelopersTeam:2018hym,Bonvini:2019wxf}
that the inclusion of small-$x$ resummation is beneficial
and improves the agreement with data,
but they also show that the QSPDF parametrization is not flexible enough at small $x$
to give a good description even when small-$x$ resummation is included.

\begin{figure}[t]
  \centering
  \includegraphics[width=0.328\textwidth,page=1]{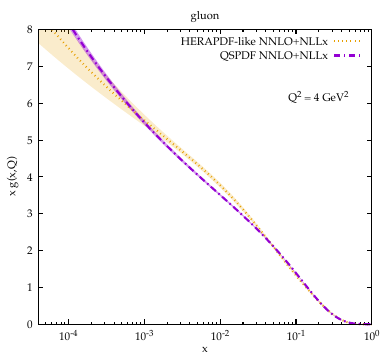}
  \includegraphics[width=0.328\textwidth,page=3]{PDF_comparison__QSPDF_HERAPDF_res.pdf}
  \includegraphics[width=0.328\textwidth,page=5]{PDF_comparison__QSPDF_HERAPDF_res.pdf}\\
  \includegraphics[width=0.328\textwidth,page=2]{PDF_comparison__QSPDF_HERAPDF_res.pdf}
  \includegraphics[width=0.328\textwidth,page=4]{PDF_comparison__QSPDF_HERAPDF_res.pdf}
  \includegraphics[width=0.328\textwidth,page=6]{PDF_comparison__QSPDF_HERAPDF_res.pdf}
  \caption{Same as figure~\ref{fig:QSPDF} but including small-$x$ resummation at NNLO+NLL$x$ accuracy.}
  \label{fig:QSPDFres}
\end{figure}

As far as PDFs are concerned, we plot in figure~\ref{fig:QSPDFres} both ``resummed'' PDF sets, using the same structure as before.
We observe that most PDFs are essentially unchanged, with small differences visible only in the $\bar u$ distribution,
with the exception of the gluon PDF, that changes significantly in the HERAPDF-like fit, getting much closer to
the QSPDF gluon, which is instead basically unchanged.
This shows that indeed the QSPDF parametrization is more suitable for fitting PDFs with all-order resummation
of small-$x$ logarithms than without.

\begin{figure}[t]
  \centering
  \includegraphics[width=0.328\textwidth,page=1]{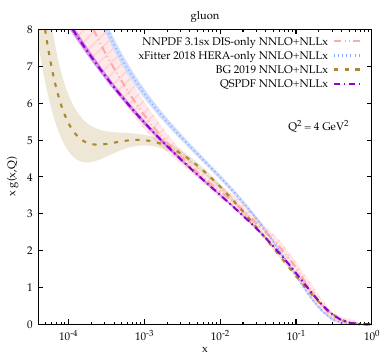}
  \includegraphics[width=0.328\textwidth,page=3]{PDF_comparison__QSPDF_others_res.pdf}
  \includegraphics[width=0.328\textwidth,page=5]{PDF_comparison__QSPDF_others_res.pdf}\\
  \includegraphics[width=0.328\textwidth,page=2]{PDF_comparison__QSPDF_others_res.pdf}
  \includegraphics[width=0.328\textwidth,page=4]{PDF_comparison__QSPDF_others_res.pdf}
  \includegraphics[width=0.328\textwidth,page=6]{PDF_comparison__QSPDF_others_res.pdf}
  \caption{Same as figure~\ref{fig:QSPDFres}, but comparing QSPDF with
    NNPDF3.1sx~\cite{Ball:2017otu} (dot-dot-dashed salmon),
    the 2018 \texttt{xFitter} low-$x$ study of Ref.~\cite{xFitterDevelopersTeam:2018hym} (dotted blue) and
    BG 2019~\cite{Bonvini:2019wxf} (dashed brown).}
  \label{fig:QSPDFothers_res}
\end{figure}

To conclude, we also compare the QSPDF fit at NNLO+NLL$x$ with analogous resummed fits
from Refs.~\cite{Ball:2017otu,xFitterDevelopersTeam:2018hym,Bonvini:2019wxf} in figure~\ref{fig:QSPDFothers_res}.
There are differences at medium-high $x$ and in the valence distributions between the sets that are due
to the parametrization and the dataset which are not very useful to compare now.
Let's focus instead on the low-$x$ region.
We see in general good agreement in the quark sea contributions, with the uncertainty
from the NNPDF3.1sx fit being large enough to cover all other curves
(except the $\bar u$ distribution from the 2018 \xfitter\ study with resummation~\cite{xFitterDevelopersTeam:2018hym}
which is slightly higher but still very close).
In the gluon PDF we see a general tendency to grow at small $x$,
with the BG 2019 fit from Ref.~\cite{Bonvini:2019wxf} having a peculiar shape that makes it different
from the other sets. This shape is due to the particular parametrization as well as the use of
a newer version of the resummation code \texttt{HELL} with respect to the previous two fits,
differing from the previous version by subleading logarithmic contributions~\cite{Bonvini:2018xvt,Bonvini:2018iwt},
as documented in Ref.~\cite{Bonvini:2019wxf} itself.
This newer version of \texttt{HELL} is the same used here, but both QSPDF and HERAPDF-like parametrizations
are not flexible enough to produce a similar shape.

As a final observation, we note that the shape of the gluon obtained with small-$x$ resummation
is similar to what is obtained with the recent MSHT (approximate) N$^3$LO fit~\cite{McGowan:2022nag}.
Indeed the small-$x$ logarithms appearing at this order behave in a way similar
to their all-order resummation, at least in a region of intermediate $x\sim 10^{-3}$,
thus providing a sort of approximation of the all-order behaviour in that region.
Performing a QSPDF fit at (approximate) N$^3$LO using \xfitter\ is however not possible at the moment
due to the lack of the necessary theoretical ingredients in the code.

\section{More flexible QSPDF parametrization}
\label{sec:flex}

In section~\ref{sec:fit} we have seen that the QSPDF parametrization is able to give a reasonable description
of the PDFs at medium-high $x$, but it is not sufficiently flexible at small $x$ to describe the data well.
In particular, the gluon PDF parametrization Eq.~\eqref{eq:gluonPDF} is very constrained and cannot
produce the variety of shapes obtained in PDF fits using more flexible parametrizations.

In this section we thus consider a minimal modification of the QSPDF parametrization
that increases the flexibility of the gluon PDF at small $x$.
To do so, we follow the suggestion of Ref.~\cite{Bonvini:2019wxf}
of using a polynomial in $\log x$ to model the shape at small $x$, and modify
the gluon parametrization Eq.~\eqref{eq:gluonPDF} as
\begin{equation}\label{eq:gluonPDFflex}
  x g(x, \mu_0^2) = A_g h_-(x;b_g,0) \[1+F_g\log x+G_g\log^2x\].
\end{equation}
The polynomial contribution in $\log x$ does not modify the behaviour of the PDF
at large $x$, where the statistical model is meant to be physically motivated,
and gives additional degrees of freedom to model the low-$x$ region,
where instead the model behaviour $x^{b_g}$ is just phenomenological.

In Eq.~\eqref{eq:gluonPDFflex}, $F_g$ and $G_g$ are two new parameters to be fitted.\footnote
{We use the same notation of Ref.~\cite{Bonvini:2019wxf} for a direct comparison.}
Moreover, we also decide to unlink $b_g$ from $\tilde b$, considering it a free parameter.
In this way we have three extra free parameters with respect to the QSPDF parametrization of section~\ref{sec:model}.
However, we have verified that with this new choice of parametrization for the gluon
it is possible to fix the value of the $\bar b$ parameter entering the antiquark parametrization
without decreasing the fit quality.
We choose as default value $\bar b=b$, and we verified that other reasonable choices (e.g.\ $\bar b=b/2$) do not change the fit quality.
According to this procedure, we have to a total of 11 free parameters to be fitted,
\beq
\xb, b, \tilde b, b_g, F_g, G_g, \tilde A, X^\uparr_u, X^\dwarr_u, X^\uparr_d, X^\dwarr_d,
\eeq
which is just two more parameters with respect to the default QSPDF parametrization,
and three parameters less than the HERAPDF parametrization.
We dub this alternative parametrization QSPDFflex.

\begin{table}
\centering
\begin{tabular}{lcc}
  & QSPDFflex &  QSPDFflex  \\
  Contribution to $\chi^2$ & (NNLO) &  (NNLO+NLL$x$)  \\
  \midrule
  subset NC $e^+$ 920      & $412/363$   & $401/363$  \\
  subset NC $e^+$ 820      & $ 71/ 68$   & $ 67/ 68$  \\
  subset NC $e^+$ 575      & $224/249$   & $220/249$  \\
  subset NC $e^+$ 460      & $220/200$   & $223/200$  \\
  subset NC $e^-$          & $226/159$   & $228/159$  \\
  subset CC $e^+$          & $ 55/ 39$   & $ 53/ 39$  \\
  subset CC $e^-$          & $ 63/ 42$   & $ 62/ 42$  \\
  correlation term + log term   & $80-17$      & $75-17$ \\
  \bf\boldmath Total $\chi^2/\rm{d.o.f.}$  &\boldmath $1334/1109$ &\boldmath $1311/1109$ \\
\end{tabular}
\caption{Same as table~\ref{tab:chi2newsetting}, showing the $\chi^2$ breakdown for QSPDFflex at NNLO and NNLO+NLL$x$.}
\label{tab:chi2QSPDFflex}
\end{table}

We start considering the fit quality of the QSPDFflex parametrization,
both at NNLO and with small-$x$ resummation at NNLO+NLL$x$.
We report the $\chi^2$ breakdown in table~\ref{tab:chi2QSPDFflex}.
We observe immediately that now the quality is comparable to the analogous fits
obtained with the HERAPDF parametrization.
As expected, the improvement is driven by a better description of the $E=920$~GeV dataset
containing the small-$x$ data.
More precisely, taking into account the different number of parameters,
the fit quality is basically identical for QSPDFflex and HERAPDF-like,
making each parametrization as good as the other.
As such, each of them represent a measure of the parametrization bias of the other,
and could be used for constructing a parametrization uncertainty.

\begin{figure}[t]
  \centering
  \includegraphics[width=0.328\textwidth,page=1]{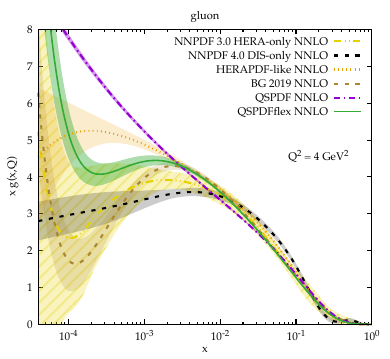}
  \includegraphics[width=0.328\textwidth,page=3]{PDF_comparison__QSPDFflex_others.pdf}
  \includegraphics[width=0.328\textwidth,page=5]{PDF_comparison__QSPDFflex_others.pdf}\\
  \includegraphics[width=0.328\textwidth,page=2]{PDF_comparison__QSPDFflex_others.pdf}
  \includegraphics[width=0.328\textwidth,page=4]{PDF_comparison__QSPDFflex_others.pdf}
  \includegraphics[width=0.328\textwidth,page=6]{PDF_comparison__QSPDFflex_others.pdf}
  \caption{Comparison of QSPDF (dot-dashed purple) and QSPDFflex (solid green) at NNLO
    with our HERAPDF-like fit (dotted orange), NNPDF3.0 HERA-only (dot-dot-dashed yellow), NNPDF4.0 DIS-only (dashed black)
    and BG (dashed brown)
    for the gluon, total singlet, $\bar u$, $\bar d+\bar s$, $u_v$ and $d_v$ PDFs.}
  \label{fig:QSPDFflexothers}
\end{figure}

We now move to comparing the PDFs.
In figure~\ref{fig:QSPDFflexothers} we show the QSPDFflex set at NNLO together with all the NNLO PDFs considered so far.
In particular, we see that gluon PDF in the new fit (solid green curve) is rather different from the
previous QSPDF gluon, as now there is a non-trivial shape that tends to reduce the size of the gluon
below $x\sim10^{-3}$ before rising again at $x\sim10^{-4}$.
This shape is similar to the one obtained in the BG fit of Ref.~\cite{Bonvini:2019wxf},
even though in that case the drop and growth are stronger, and also close to the NNPDF3.0 HERA-only fit.
More in general, all PDF parametrizations except the QSPDF one predict a gluon that either flattens
or decreases before possibly growing again at small $x$.
As far as the other PDFs are concerned, the difference between the QSPDFflex and the QSPDF sets
is very small or totally negligible, as expected given that the biggest change is in the parametrization of the gluon.

\begin{figure}[t]
  \centering
  \includegraphics[width=0.328\textwidth,page=1]{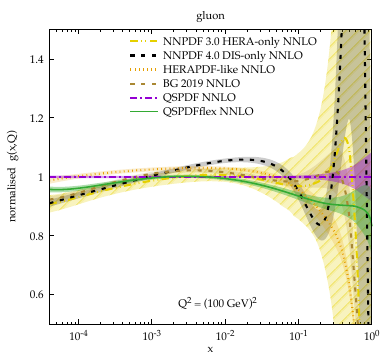}
  \includegraphics[width=0.328\textwidth,page=3]{PDF_comparison_ratio__QSPDFflex_others.pdf}
  \includegraphics[width=0.328\textwidth,page=5]{PDF_comparison_ratio__QSPDFflex_others.pdf}\\
  \includegraphics[width=0.328\textwidth,page=2]{PDF_comparison_ratio__QSPDFflex_others.pdf}
  \includegraphics[width=0.328\textwidth,page=4]{PDF_comparison_ratio__QSPDFflex_others.pdf}
  \includegraphics[width=0.328\textwidth,page=6]{PDF_comparison_ratio__QSPDFflex_others.pdf}
  \caption{Same as figure~\ref{fig:QSPDFflexothers} but for $Q=100$~GeV, showing the various curves as ratios to QSPDF.}
  \label{fig:QSPDFflexothers100}
\end{figure}

We also consider in figure~\ref{fig:QSPDFflexothers100} the same PDFs at the electroweak scale $Q=100$~GeV,
plotted as a ratio to QSPDFs.
For gluon and anti-quark PDFs we see a generally good agreement between the various sets in the small-$x$ region,
with differences at most at the 10\% level, even though not always within the uncertainty.
This shows that DGLAP evolution reduces the discrepancies between the sets, especially in the gluon PDF.
At larger $x\gtrsim0.1$ the uncertainties get larger as well as the differences between sets,
due to the lack of direct experimental constraints this region,
which is thus strongly dependent on the functional forms adopted.
The valence distributions are characterised by larger relative uncertainties and differences
(note also the larger range shown in the plot), especially in the low-$x$ region.
In particular, we observe that the HERAPDF-like fit predicts a very different $u_v$ distribution from all the other sets.
Similarly, the $d_v$ distribution of the HERAPDF-like set is very different from QSPDF(flex) and NNPDF,
but almost identical to the BG set.
The large uncertainties and differences at large $x$ also shows that
the use of non-vanishing functions in $x=1$, Eq.~\eqref{eq:QSx=1},
in place of a more standard power suppression in $1-x$,
does not represent an issue when describing the data.

\begin{figure}[t]
  \centering
  \includegraphics[width=0.328\textwidth,page=1]{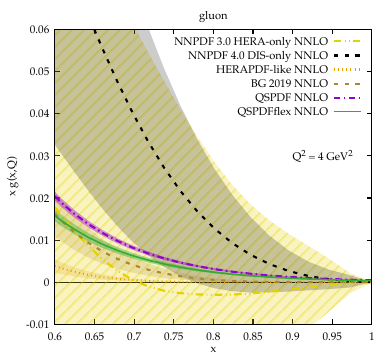}
  \includegraphics[width=0.328\textwidth,page=3]{PDF_comparison__QSPDFflex_others_largex.pdf}
  \includegraphics[width=0.328\textwidth,page=5]{PDF_comparison__QSPDFflex_others_largex.pdf}\\
  \includegraphics[width=0.328\textwidth,page=2]{PDF_comparison__QSPDFflex_others_largex.pdf}
  \includegraphics[width=0.328\textwidth,page=4]{PDF_comparison__QSPDFflex_others_largex.pdf}
  \includegraphics[width=0.328\textwidth,page=6]{PDF_comparison__QSPDFflex_others_largex.pdf}
  \caption{Same as figure~\ref{fig:QSPDFflexothers} zoomed in at large $x$.}
  \label{fig:QSPDFflexotherslargex}
\end{figure}

To further verify the impact of using the functional form
based on Fermi-Dirac and Bose-Einstein distributions,
we plot in figure~\ref{fig:QSPDFflexotherslargex} a zoom of the large-$x$ region
of figure~\ref{fig:QSPDFflexothers}.
We can appreciate that gluon and anti-quark distributions are essentially indistinguishable
from other sets where PDFs vanish in $x=1$,
and are compatible with them within uncertainty in the $x\to1$ region.
Quark PDFs are instead more sensitive to the functional form.
In particular, the valence distributions of QSPDF and QSPDFflex
tend to be higher than the HERAPDF-like fit for $x\gtrsim0.7$,
staying visibly larger than zero at $x=1$.
This difference is inherited by the total singlet PDF.
We note however that QSPDF and QSPDFflex are fully within the NNPDF3.0 uncertainty band
even at very high $x$, due to the large uncertainty bands of the set
in turn due to the lack of data in this region.
Note that the NNPDF4.0 set has a much reduced uncertainty.
This is partly due to the larger dataset, that includes several DIS data at rather high-$x$
(e.g.\ BCDMS~\cite{BCDMS:1989qop} reaches $x=0.75$,
NMC~\cite{NewMuon:1996uwk} reaches $x=0.68$,
CHORUS~\cite{CHORUS:2005cpn} reaches $x=0.65$),
and partly also to the improved fitting methodology~\cite{NNPDF:2021njg}
that favours smoother replicas allowing for smaller uncertainties in the extrapolation region $x\gtrsim0.75$.
Since NNPDF4.0 explicitly assumes a power behaviour $(1-x)^\alpha$ vanishing at large $x$ ($\alpha>0$),
the small uncertainty in the extrapolation region $x\gtrsim0.75$ is not necessarily reliable.
Therefore, the inconsistency of our quark PDFs with NNPDF4.0 in this region is not
indicative of a preference towards a stronger large-$x$ suppression.

\begin{figure}[t]
  \centering
  \includegraphics[width=0.328\textwidth,page=1]{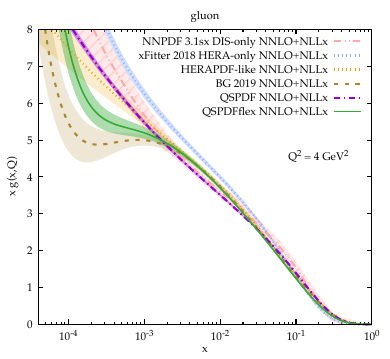}
  \includegraphics[width=0.328\textwidth,page=3]{PDF_comparison__QSPDFflex_others_res.pdf}
  \includegraphics[width=0.328\textwidth,page=5]{PDF_comparison__QSPDFflex_others_res.pdf}\\
  \includegraphics[width=0.328\textwidth,page=2]{PDF_comparison__QSPDFflex_others_res.pdf}
  \includegraphics[width=0.328\textwidth,page=4]{PDF_comparison__QSPDFflex_others_res.pdf}
  \includegraphics[width=0.328\textwidth,page=6]{PDF_comparison__QSPDFflex_others_res.pdf}
  \caption{Comparison of QSPDF (dot-dashed purple) and QSPDFflex (solid green) at NNLO+NLL$x$
    with our HERAPDF-like fit (dotted orange), NNPDF3.1sx DIS-only (dot-dot-dashed salmon), xFitter 2018 (dotted blue)
    and BG (dashed brown)
    for the gluon, total singlet, $\bar u$, $\bar d+\bar s$, $u_v$ and $d_v$ PDFs.}
  \label{fig:QSPDFflexothers_res}
\end{figure}

When including the resummation of small-$x$ logarithms, the agreement between the various curves improves.
This is shown in figure~\ref{fig:QSPDFflexothers_res}, again comparing the new QSPDFflex set with all the
PDF sets with resummation considered before.
Again, the gluon is the one that exhibits the biggest difference between QSPDF and QSPDFflex,
but now the QSPDFflex gluon grows as the QSPDF one, but with some oscillations that resemble
those of the BG set~\cite{Bonvini:2019wxf}.
We keep seeing that the QSPDFflex is close to QSPDF for the other PDFs,
but now there is a more marked difference in the sea quarks and in the $d_v$ distribution at medium $x$.

\begin{figure}[t]
  \centering
  \includegraphics[width=0.328\textwidth,page=1]{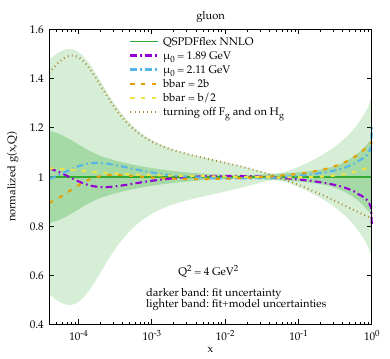}
  \includegraphics[width=0.328\textwidth,page=3]{PDF_comparison__QSPDF_model_unc_rat.pdf}
  \includegraphics[width=0.328\textwidth,page=5]{PDF_comparison__QSPDF_model_unc_rat.pdf}\\
  \includegraphics[width=0.328\textwidth,page=2]{PDF_comparison__QSPDF_model_unc_rat.pdf}
  \includegraphics[width=0.328\textwidth,page=4]{PDF_comparison__QSPDF_model_unc_rat.pdf}
  \includegraphics[width=0.328\textwidth,page=6]{PDF_comparison__QSPDF_model_unc_rat.pdf}
  \caption{Model uncertainties at NNLO obtained by various variations as indicated in the text.
    The plots are presented as ratios to the central PDF,
    for the gluon, total singlet, $\bar u$, $\bar d+\bar s$, $u_v$ and $d_v$ PDFs.}
  \label{fig:QSPDFflex_model_unc_rat}
\end{figure}

We now discuss the uncertainties. We observe that the QSPDFflex has similar (small) uncertainties to QSPDF,
except for the gluon which has a larger band at small $x$, due to the presence of extra parameters.
For the more promising QSPDFflex parametrization we also consider model uncertainties, in particular parametrization uncertainty.
Specifically, we investigate the effect of changing the initial scale of the parametrization,
raising\footnote
{When we raise the initial scale, we also have to raise the charm threshold to keep the condition $\mu_c>\mu_0$.
  The contribution of raising $\mu_c$ is less important than that of raising $\mu_0$,
  and indeed the effect is rather symmetric to the lower variation of $\mu_0$ where $\mu_c$ is kept fixed.}
or lowering it by 0.11~GeV,
and of modifying parameters or functional form:
we vary $\bar b$ up and down to $\bar b=2b$ and $\bar b=b/2$,
and replace the linear logarithmic gluon term $F_g\log x$
with a cubic term $H_g\log^3x$ (following an analogous variation performed in Ref.~\cite{Bonvini:2019wxf}).
The relative effect of each individual variation is shown at NNLO in
figure~\ref{fig:QSPDFflex_model_unc_rat}
(the relative uncertainties at NNLO+NLL$x$ are similar).
We notice that the gluon uncertainty is dominated at medium-small $x$ by the effect of
changing the parametrization with a cubic logarithmic terms in place of the linear one, as expected.
Its effect on other PDFs is mild, and concentrated at high $x$, indirectly induced by the momentum sum rule.
The $\bar b$ variations have larger effects on the (anti)quarks, and in particular on the valence distributions
at medium-low $x$, most importantly for the $d_v$.
The variation of the fit scale $\mu_0$ are mild and similar in all PDFs, giving bigger effects at
small and high $x$.

\begin{figure}[t]
  \centering
  \includegraphics[width=0.328\textwidth,page=1]{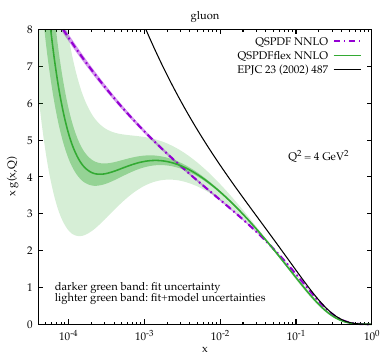}
  \includegraphics[width=0.328\textwidth,page=3]{PDF_comparison__Buccella.pdf}
  \includegraphics[width=0.328\textwidth,page=5]{PDF_comparison__Buccella.pdf}\\
  \includegraphics[width=0.328\textwidth,page=2]{PDF_comparison__Buccella.pdf}
  \includegraphics[width=0.328\textwidth,page=4]{PDF_comparison__Buccella.pdf}
  \includegraphics[width=0.328\textwidth,page=6]{PDF_comparison__Buccella.pdf}
  \caption{Comparison of QSPDFflex (solid green), showing also the total (fit+model) uncertainty,
    with QSPDF (dot-dashed purple) and the older determination of Ref.~\cite{Bourrely:2001du} (solid black, without uncertainties),
    for the gluon, total singlet, $\bar u$, $\bar d+\bar s$, $u_v$ and $d_v$ PDFs.}
  \label{fig:Buccella}
\end{figure}

To conclude the section, we show the actual PDFs of the QSPDFflex set including the model uncertainties in Fig.~\ref{fig:Buccella}.
We also show the QSPDF set for comparison.
The lighter-green uncertainty band represents the total (fit+model) uncertainty,
obtained by summing in quadrature the fit uncertainty (corresponding to the darker-green band)
and each individual model variation with respect to the central PDF.
We note that the gluon and $d_v$ distributions have visibly larger total bands,
while for the others the difference is less marked, but still visible in some regions (especially low $x$ and medium-high $x$).

\section{Implications}
\label{sec:implications}

Having established the possibility of fitting HERA data with reasonable/good quality with
the QSPDF and QSPDFflex parametrizations, we now want to comment on some possible physical implications of this study.

\subsection{Comparison with previous model determinations}

First of all, we may consider the success in fitting the data as a success of the statistical model behind the QSPDF parametrization.
It is true that the original model of Ref.~\cite{Bourrely:2001du} leading to the QSPDF parametrization
does not allow to fit the data with high quality, but still the quality is reasonable
and moreover we have seen that most of the problems come from the limited flexibility of the gluon at small $x$.
Admittedly, the statistical model of Ref.~\cite{Bourrely:2001du} is designed
to describe the high-$x$ region of the PDFs, and the ingredients needed to describe
the small $x$ region, i.e.\ the $x^b$ factors and the ``diffractive'' (sea) quark terms,
are only phenomenological.
Therefore, we believe that the QSPDFflex parametrization of section~\ref{sec:flex}
is still in agreement with the original ideas of the statistical model,
as in particular it does not change the description of the large-$x$ PDFs.
In this sense, the better description achieved with the QSPDFflex parametrization
can be considered as a success of the statistical model.

As we mentioned before, the model is agnostic about QCD interactions and therefore it does
not know about the perturbative order and the factorization scheme and scale dependence.
In this sense, the good description of the data may seem surprising.
However, we have to recall that we are using the model as a provider for a parametrization,
and it is the parametrization that works well.
We can also guess why it is so.
Essentially, the model assumes the original Feynman's parton model,
i.e.\ the LO approximation of the QCD factorization theorem.
It is well known that LO PDFs have shapes that are similar to $\MSbar$-scheme NLO and NNLO PDFs at large $x$,
and so it is likely that the same parametrization, with different parameters,
is able to describe also NNLO PDFs.
At small $x$ differences are more marked between various orders, and indeed we had to
modify the parametrization of the gluon PDF in the QSPDFflex set to obtain a reasonable description.

\begin{table}
  \centering
  {\small
\begin{tabular}{lccccc}
  Fitted& QSPDF & QSPDF & QSPDFflex & QSPDFflex & Ref.~\cite{Bourrely:2001du} \\
  param. & NNLO & NNLO+NLL$x$ & NNLO & NNLO+NLL$x$ & \\
  \midrule
  $\xb$        & $0.0950\pm0.0011$ & $0.0932\pm0.0012$ & $0.0964\pm0.0017$ & $0.0971\pm0.0018$ & 0.09907 \\
  $b$          & $0.557\pm0.009$ & $0.54\pm0.03$ & $0.538\pm0.012$ & $0.545\pm0.012$ & 0.40962 \\
  $\bar b$     & $0.00016\pm0.00003$ & $0.000019\pm0.000003$ & $b$ & $b$ & $2b$ \\
  $\tilde b$   & $-0.1700\pm0.0018$ & $-0.172\pm0.003$ & $-0.179\pm0.005$ & $-0.169\pm0.005$ & $-0.25347$ \\
  $b_g$        & $\tilde b+1$ & $\tilde b+1$ & $0.440\pm0.013$ & $0.434\pm0.016$ & $\tilde b+1$ \\
  $F_g$        & - & - & $0.212\pm0.004$ & $0.207\pm0.004$ & - \\
  $G_g$        & - & - & $0.0116\pm0.0004$ & $0.0111\pm0.0004$ & - \\
  $\tilde A$   & $0.159\pm0.003$ & $0.151\pm0.004$ & $0.146\pm0.006$ & $0.156\pm0.006$ & 0.08318 \\
  $X^\uparr_u$ & $0.410\pm0.006$ & $0.415\pm0.007$ & $0.407\pm0.008$ & $0.404\pm0.008$ & 0.46128 \\
  $X^\dwarr_u$ & $0.21\pm0.02$ & $0.216\pm0.019$ & $0.18\pm0.03$ & $0.18\pm0.03$ & 0.29766 \\
  $X^\uparr_d$ & $0.11\pm0.03$ & $0.13\pm0.03$ & $0.13\pm0.02$ & $0.12\pm0.03$ & 0.22775 \\
  $X^\dwarr_d$ & $0.292\pm0.008$ & $0.292\pm0.009$ & $0.276\pm0.008$ & $0.278\pm0.008$ & 0.30174 \\
  \midrule
  $A$      & 3.4 & 3.2 & 3.6 & 3.6 & 1.74938 \\
  $\bar A$ & 0.00002 & 0.00006 & 0.13 & 0.08 & 1.90801 \\
  $A_g$    & 17 & 18 & 11 & 11 & 14.27535
\end{tabular}
}
\caption{Values of the fitted parameters (including uncertainties) for the QSPDF and QSPDFflex sets, both at NNLO and NNLO+NLL$x$.
  In the last three lines we also provide the values of the additional normalization parameters that are determines from the sum rules
  (indicating the central value only).
  In the last column we also report the values of the parameters from the determination of Ref.~\cite{Bourrely:2001du},
  which were given without uncertainty and with the same number of digits shown here.}
\label{tab:pars}
\end{table}

It is interesting to compare the values of the model parameters with previous determinations.
In table~\ref{tab:pars} we list the values of the parameters for the four fits considered,
namely QSPDF and QSPDFflex, both at NNLO and NNLO+NLL$x$, as well as those from Ref.~\cite{Bourrely:2001du}.
We immediately notice that the values of the parameters is very similar across all our fits.
The only exception is $\bar b$, which is fixed to be equal to $b$ in the QSPDFflex fits,
whose value is very different from the value found in the QSPDF fits.
In fact, the very small value of the $\bar b$ parameter in the QSPDF fits,
which describes a strong small-$x$ growth of the antiquark valence component,
is compensated by a very small value of $\bar A$ computed by the sum rules
(see appendix~\ref{sec:sumrules}), making the effect of that term very small.

Comparing our numbers with the values obtained in Ref.~\cite{Bourrely:2001du}
we see that there is generally a good agreement between the parameters.
In particular, the ``effective temperature'' $\bar x$, which governs the large-$x$ drop of all PDFs,
is very stable across the various determinations.
The quark ``potentials'' are in good agreement,
but they are all smaller in our fits, in particular $X_d^\uparr$ and to a lesser extent $X_u^\dwarr$.
This difference is likely due to the absence in our determination of information from polarized distributions.
The $b$ parameter governing the small-$x$ drop of the valence distributions is always slightly larger in our fits,
as well as the $\tilde b$ parameter governing the small-$x$ growth of sea quarks,
which in turn determines a larger $\tilde A$ coefficient to compensate.
In the last three lines of the table the values of the parameters $A,\bar A,A_g$,
determined from the sum rules, are also shown.
As the value of $\bar A$ strongly depends on the value of $\bar b$,
it is very different in the various families of fits.

The effect of these different parameters is shown in figure~\ref{fig:Buccella},
where the PDFs from Ref.~\cite{Bourrely:2001du} are compared with our QSPDF(flex) NNLO sets.
We see clear distortions in all distributions between the old and new sets.
At high $x$ all distributions behave in the same way, as this region is predominantly
governed by the $\bar x$ parameter which is almost the same for all PDFs and secondarily by the
quark potentials $X_q^\uda$ which are similar.
The PDFs of Ref.~\cite{Bourrely:2001du} are all harder at small $x$, including the valence distributions,
and compensate this with smaller quark sea distributions at medium $x$ and with smaller valence peaks.
These differences are certainly due to the fact that Ref.~\cite{Bourrely:2001du}
fits a bunch of DIS data from different experiments, including polarized data,
all at a $Q^2$ scale close to $\mu_0^2=4$~GeV$^2$, thus containing different information with respect to our dataset.
As a comparison, we have computed the $\chi^2$ of the PDF set of Ref.~\cite{Bourrely:2001du} with our fit setting,
finding more than 6000 units, showing that this PDF set is very far from giving an acceptable description of HERA data.

\subsection{The anti-up anti-down asymmetry in the proton}

An important implication of the statistical model is that the difference $\bar d-\bar u$ is greater than zero.
A positive value for the first Mellin moment of this difference
was first determined by the NMC experiment~\cite{NewMuon:1991hlj},
which found a defect in the Gottfried sum rule~\cite{Gottfried:1967kk}.
This confirmed the conjecture by Niegawa and Sasaki~\cite{Niegawa:1974hk} and
by Feynman and Field~\cite{Field:1976ve} that, as a consequence of Pauli principle,
in the proton there are more $\bar{d}$ than $\bar{u}$.

The positivity of $\bar d-\bar u$ is a consequence of the values of the potentials $X^\uda_{u,d}$
that we can understand analytically.
Looking at Eq.~\eqref{eq:PDFparqv} we note that, for each polarization, there is one term
that dominates over the other. Using explicitly Eq.~\eqref{eq:CudX} and grouping terms with the same potential,
\begin{align}
  x q_v (x, \mu_0^2) & =  A X_q^\uparr h_+(x; b, X_q^\uparr) - \frac{\Ab}{X_q^\uparr} h_+(x;\bb, -X_q^{\uparr}) \nonumber\\
                    & +  A X_q^\dwarr h_+(x; b, X_q^\dwarr) - \frac{\Ab}{X_q^\dwarr} h_+(x;\bb, -X_q^{\dwarr}),
\end{align}
we see that, for positive potentials $X^\uda_q>0$ as given by the fit,
the second term proportional to $\bar A$ in each line is exponentially suppressed with respect to each first term
by a factor $\exp(-2X^\uda_q/\bar x)$ at medium/large $x$.
Therefore, the size of the valence distribution is dominated by the $A$ terms,
and in particular by the polarization component characterised by the largest potential,
which is $\max(X^\uparr_u, X^\dwarr_u)=X^\uparr_u$ for the up quark
and $\max(X^\uparr_d, X^\dwarr_d)=X^\dwarr_d$ for the down quark\footnote
{Note that since the parametrization of quarks is simply the sum of the two independent polarization components,
  differing only by the potentials, there is a symmetry in the parametrizations for the
  exchange $X^\uparr_q \leftrightarrow X^\dwarr_q$.
  Therefore, which potential is larger is just a matter choice, i.e.\ the fit can equivalently
  find a minimum or its symmetric where the potentials are swapped.
  The hierarchy shown here reproduces the hierarchy found in the original publications
  where additional information from polarized data were included.}
(table~\ref{tab:pars}).
Moreover, since the valence distribution of the up quark is larger (roughly by a factor of two)
than the valence distribution for the down, it follows that $\max(X^\uparr_u, X^\dwarr_u)>\max(X^\uparr_d, X^\dwarr_d)$,
namely $X^\uparr_u>X^\dwarr_d$, which is indeed verified in all fits.
From the fit results we also see that the potential for the subdominant polarizations
is always larger for the up quark than for the down quark, $X^\dwarr_u>X^\uparr_d$.
While this is not directly related to striking features of the parametrization,
it is a property that ensures that the shapes of the valence distributions are in agreement with data
and with the sum rules.\footnote
{We were not able to prove that this inequality has to be satisfied based on strict conditions.
  Therefore, we can only say that it is very unlikely that a good fit can violate this hierarchy,
  and indeed all fits so far led to values satisfying this hierarchy.}
This inequality immediately implies that $\bar d>\bar u$. Indeed their difference is given by
\begin{align}\label{eq:dubar}
  x \bar d (x, \mu_0^2) - x \bar u (x, \mu_0^2)
  & =  \bar A \bigg[ \frac1{X_d^\uparr} h_+(x;\bb, -X_d^{\uparr}) + \frac1{X_d^\dwarr} h_+(x;\bb, -X_d^{\dwarr}) \nonumber\\
  & \qquad- \frac1{X_u^\uparr} h_+(x;\bb, -X_u^{\uparr}) - \frac1{X_u^\dwarr} h_+(x;\bb, -X_u^{\dwarr}) \bigg],
\end{align}
which is dominated by the smallest potentials, and it is positive if
$\min(X^\uparr_u, X^\dwarr_u)>\min(X^\uparr_d, X^\dwarr_d)$, namely $X^\dwarr_u>X^\uparr_d$,
which is exactly the condition mentioned before.\footnote
{This conclusion assumes that $\bar A>0$. This is always the case in
  the fits considered here, but it may not be the case in general.
  However, $\bar A>0$ ensures that the anti-quark distributions are positive in the high-$x$ region,
  so we find it reasonable to assume that this will always be the case in good fits.}

\begin{figure}[t]
  \centering
  \includegraphics[width=0.49\textwidth,page=1]{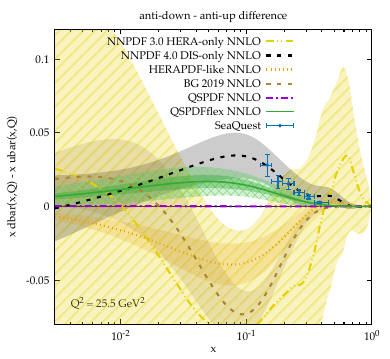}
  \includegraphics[width=0.49\textwidth,page=1]{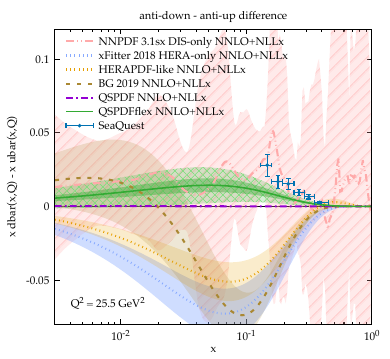}
  \caption{The difference $\bar d-\bar u$ in all the PDF sets considered so far, at NNLO (left) and NNLO+NLL$x$ (right).
    For the QSPDFflex result the total uncertainty band is also shown in lighter green with a crossed pattern.
    The scale of the plot is $Q^2=25.5$~GeV$^2$ to match that of the SeaQuest data points~\cite{SeaQuest:2021zxb,FNALE906:2022xdu}
    also shown in the plot.}
  \label{fig:dubar}
\end{figure}

The numerical results are reported in figure \ref{fig:dubar} for several PDF sets
at fixed order (left plot) and with small-$x$ resummation (right plot).
We notice that all PDF sets obtained using only HERA data tend to
predict a negative $\bar d-\bar u$ difference in the valence region,
with the exception of the QSPDF and QSPDFflex sets.
The inclusion of data sensitive to quark flavours, e.g.\ charged current data from DIS experiments
as in the NNPDF4.0 DIS-only set,
twists the situation by predicting a positive $\bar d-\bar u$ difference in the valence region.
This is further confirmed in global PDF fits (see e.g.\ Ref.~\cite{ATLAS:2021vod}).
We thus conclude that HERA data alone not only are not able to predict the flavour separation,
but they also tend to be in better agreement with a negative $\bar d-\bar u$ difference irrespective of the
parametrization used, again with the exception of QSPDFs.

Now let us focus on our QSPDF and QSPDFflex fits.
We observe that the latter gives a positive $\bar d-\bar u$ distribution,
in agreement with the discussion above about the values of the potentials.
However, the QSPDF predicts essentially the same values for $\bar d$ and $\bar u$,
giving a vanishing difference.
This effect is the result of the very small value of $\bar b$ coming from the fit
that forces $\bar A$ to be almost zero. As the difference $\bar d-\bar u$
is proportional to $\bar A$, Eq.~\eqref{eq:dubar}, it becomes consequently very small.
We suspect that the small value of $\bar b$ found by the fit is driven by this effect:
HERA data favours a negative $\bar d-\bar u$, and the fit finds the value of the parameters
that makes it as close as possible to negative, i.e.\ zero.
Conversely, in the QSPDFflex parametrization where $\bar b$ is fixed to $b$,
this flexibility is no longer present and the difference $\bar d-\bar u$ remains significantly positive.

One may wonder why in the QSPDFflex parametrization we are allowed to fix $\bar b=b$
while in the QSPDF parametrization this leads to a sizeable increase in $\chi^2$.
We suspect that this is due to the greater flexibility of the gluon parametrization,
which is less linked to the diffractive (sea) term, which is then more flexible
and can thus better describe the antiquark distributions without the need for an extra degree of freedom.
We may also guess that when including data sensitive to quark flavours
leaving $\bar b$ as a free parameter can still predict a positive $\bar d - \bar u$ difference,
this time induced by the data and not by a parametrization bias.
To see this, we also report in the figure the recent data
from the SeaQuest collaboration~\cite{SeaQuest:2021zxb,FNALE906:2022xdu}.
We see that they scale in reasonable agreement with QSPDFflex, but they are higher,
in closer agreement with the NNPDF4.0 DIS-only set (despite the fact that it does not include them).
We will investigate in section~\ref{sec:newdata} the stability of this result
upon inclusion of additional data.

\subsection{Additional data}
\label{sec:newdata}

In this section we investigate the possibility of including additional data in the fit,
with emphasis on the large-$x$ behaviour of PDFs and in particular on the $\bar d-\bar u$ difference.
Unfortunately, we are limited by the datasets available in \xfitter\
(implementing new dataset ourselves would require a significant amount of work
which is far beyond the scope of this article).
As we are interested in the large-$x$ region,
we identified two datasets that are relevant for us:
old fixed-target Drell-Yan from E866~\cite{NuSea:2001idv} (39 datapoints)
and Tevatron CDF and D0 $Z$ rapidity distributions~\cite{CDF:2010vek,D0:2007djv} (56 datapoints).
In particular, the former data are given as the ratio of proton-deuteron cross section
over proton-proton cross section, giving direct access to the up/down antiquark asymmetry.

We have performed fits to HERA+E866, HERA+Tevatron and HERA+E866+Tevatron,
using both QSPDFflex and HERAPDF-like parametrization.
Unfortunately, the theoretical description of these data in \xfitter\ is limited to NLO,
so the fits use inconsistently NNLO theory for DIS and DGLAP evolution and NLO for Drell-Yan.\footnote
{We note however that the lack of NNLO corrections should be less significant for E866,
  as the data are given as a ratio of cross sections which is less sensitive to perturbative corrections.}
We must thus expect some increase in the reduced $\chi^2$ that we indeed see with both parametrizations.
For QSPDFflex, the $\chi^2/$d.o.f\ increases from the HERA-only value of 1.20
to 1.21 for HERA+Tevatron and 1.25 for HERA+E866 and HERA+E866+Tevatron.
For HERAPDF-like parametrization, the $\chi^2/$d.o.f\ increases from the HERA-only value of 1.21
to 1.26 for HERA+E866 and 1.25 for HERA+E866+Tevatron, while it decreases slightly to 1.20 for HERA+Tevatron.
The fact that these variations are very similar among the two parametrizations shows
that the QSPDFflex parametrization does a good job in fitting these data as well as HERA data.

\begin{figure}[t]
  \centering
  \includegraphics[width=0.49\textwidth,page=8]{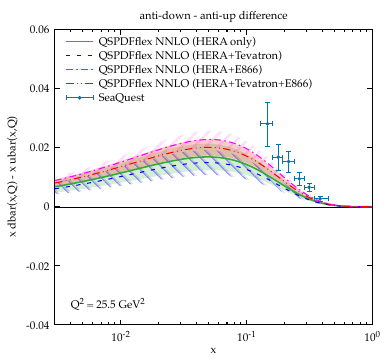}
  \includegraphics[width=0.49\textwidth,page=8]{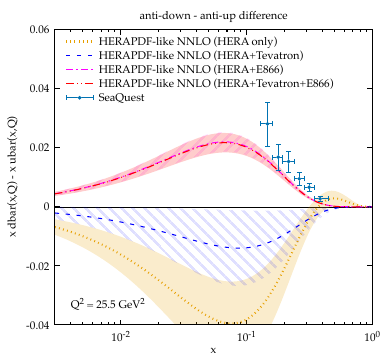}
  \caption{The difference $\bar d-\bar u$ in QSPDFflex (left) and HERAPDF-like (right) fits
    to different datasets.
    The scale of the plot is $Q^2=25.5$~GeV$^2$ to match that of the SeaQuest data points~\cite{SeaQuest:2021zxb,FNALE906:2022xdu}
    also shown in the plot.}
  \label{fig:moredata}
\end{figure}

As far as PDFs are concerned, the effect of the new data is essentially negligible for QSPDFflex,
while there are some differences for the HERAPDF-like fits, especially at high $x$.
In particular, we focus on the $\bar d-\bar u$ difference, shown in figure~\ref{fig:moredata}.
With the QSPDFflex parametrization, the inclusion of Tevatron data tends to lower slightly
the prediction, while in contrast E866 leads to a larger asymmetry, pushing the PDFs closer
to the SeaQuest data also shown in the plot.
When both datasets are included, the prediction is larger than the HERA-only fit,
but still very close to it.
Overall, the four predictions are in good agreement and we can conclude that QSPDFflex is stable upon
inclusion of different datasets, also confirming that the positive value of $\bar d-\bar u$
is a feature of the parametrization.

With the HERAPDF-like parametrization, the inclusion of Tevatron data brings the prediction
for $\bar d- \bar u$ closer to zero, but still negative in similarity with the HERA-only fit
and with similarly large uncertainty.
Once E866 data are included, the prediction becomes positive and with smaller uncertainty,
and it is close to the SeaQuest datapoints.
Moreover, it is stable upon inclusion of Tevatron data.
This shows that the HERAPDF parametrization is more flexible at high $x$
and it leads to inaccurate results when data are not sufficiently constraining,
while it provides a stable result once constraining data are included.
This behaviour is what is usually expected from a fit with unbiased parametrization.

We observe that the HERAPDF-like result with E866 data is very similar to the analogous QSPDFflex result.
This means that the QSPDFflex is also accurate, with the difference that it was so also before the inclusion of E866 data.
This is the effect of the (physically motivated) bias present in this parametrization.

\subsection{Polarized PDFs}
\label{sec:pol}

As already stressed, the parametrization of the quark PDFs is made in terms
of contributions from the individual quark polarizations.
This means that the same parameters would in principle allow to determine also polarized PDFs
within the statistical model.
Using Eq.~\eqref{eq:param-q}, the parametrizations for polarized quark and antiquark PDFs are given by
\begin{subequations}\label{eq:PDFpol}
\begin{align}
  x \Delta q (x, \mu_0^2)
  &\equiv xq^\uparr (x, \mu_0^2) - xq^\dwarr (x, \mu_0^2) \nonumber\\
  &=  A\[X_q^\uparr h_+(x; b, X_q^\uparr) - X_q^{\dwarr} h_+(x; b, X_q^{\dwarr})\], \\
  x \Delta \bar q (x, \mu_0^2)
  &\equiv x\qb^\uparr (x, \mu_0^2) - x\qb^\dwarr (x, \mu_0^2) \nonumber\\
  &= \Ab \[\frac1{X_q^\dwarr} h_+(x;\bb, -X_q^{\dwarr}) - \frac1{X_q^\uparr} h_+(x; \bb, -X_q^\uparr)\],
\end{align}
\end{subequations}
while the model assumes no polarization for gluons, i.e.\ $\Delta g=0$
(we will comment on this assumption later in this section).

Of course, a fit of unpolarized PDFs is not able to distinguish the individual polarizations.
However, if polarized data are included in the fit, it becomes possible to simultaneously 
determine polarized and unpolarized PDFs without changing the parametrization (i.e., without adding extra parameters).
The agreement (see figure~\ref{fig:Buccella}) at $x\gtrsim0.4$ with the valence distributions found in \cite{Bourrely:2001du},
where the information from polarized scattering was well described,
lets us reasonably think that the statistical model may be able to describe both unpolarized and polarized
distributions in a satisfactory way.
Testing this in practice requires quite some work, as the current \xfitter\ infrastructure
should be modified to introduce this possibility, and we therefore leave this task to future work.
Here we perform a simpler test.

\begin{figure}[t]
  \centering
  \includegraphics[width=0.328\textwidth,page=1]{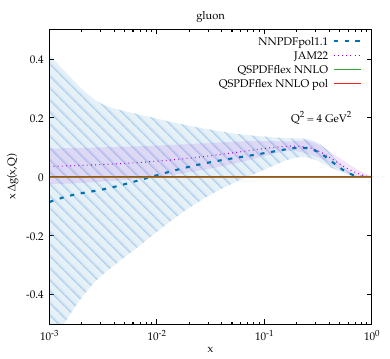}
  \includegraphics[width=0.328\textwidth,page=7]{PDF_comparison__pol.pdf}
  \includegraphics[width=0.328\textwidth,page=8]{PDF_comparison__pol.pdf}
  \\
  \includegraphics[width=0.328\textwidth,page=2]{PDF_comparison__pol.pdf}
  \includegraphics[width=0.328\textwidth,page=5]{PDF_comparison__pol.pdf}
  \includegraphics[width=0.328\textwidth,page=6]{PDF_comparison__pol.pdf}
  \caption{Comparison of polarized PDFs obtained from our QSPDFflex NNLO fit (with and without imposing the triplet sum rule)
    and the NNPDFpol1.0~\cite{Ball:2013lla} and NNPDFpol1.1~\cite{Nocera:2014gqa} sets.}
  \label{fig:pol}
\end{figure}

First of all, we consider the QSPDFflex set fitted from unpolarized data
and plot polarized PDFs Eq.~\eqref{eq:PDFpol} obtained with those parameters out of the box.
As stressed, we do not expect to find agreement with direct determinations from polarized data~\cite
{deFlorian:2009vb, Ball:2013lla, deFlorian:2014yva, Nocera:2014gqa, Ethier:2017zbq, DeFlorian:2019xxt, Zhou:2022wzm, Cocuzza:2022jye, Karpie:2023nyg, Hekhorn:2024tqm},
but we want to understand how far we are.
Therefore, in figure~\ref{fig:pol} we show the polarized PDFs constructed using Eqs.~\eqref{eq:PDFpol}
from our QSPDFflex NNLO unpolarized fit, compared with NNPDFpol1.1~\cite{Nocera:2014gqa}
and JAM22~\cite{Cocuzza:2022jye}.\footnote
{We have considered the latest version of these two families of polarized PDF fits.
  To our knowledge, the DSSV polarized PDFs~\cite{deFlorian:2009vb, deFlorian:2014yva, DeFlorian:2019xxt} are not publicly available
  in LHAPDF format.
  In any case, the PDFs shown are good representative of polarized PDFs, as for instance the most recent DSSV determination
  is in good agreement~\cite{DeFlorian:2019xxt} with NNPDFpol1.1.
  Note also that for JAM22 we have considered only the positive gluon solution, as the negative one strongly violates the
  positivity bound $|\Delta g|\leq g$, see the discussion in Ref.~\cite{deFlorian:2024utd}.}
Specifically, we plot $\Delta u$, $\Delta \bar u$, $\Delta d$, $\Delta \bar d$
and also the triplet combination
\beq\label{eq:triplet}
\Delta T_3 = \Delta u + \Delta \bar u - \Delta d - \Delta \bar d
\eeq
at the fit scale $Q^2=4$GeV$^2$. We also show the gluon $\Delta g$ for completeness.

We observe that the QSPDFflex anti-quark polarized PDFs are perfectly compatible with NNPDFpol1.1
within the (large) uncertainty of the latter.
They are also close to the JAM22 determination, except in the region $0.1\lesssim x\lesssim0.4$
where JAM22 is slightly larger (in absolute value).
The quark polarized PDFs instead are not in agreement, as $\Delta u$, $\Delta d$ and $\Delta T_3$ are all larger
(in absolute value) than their NNPDFpol1.1 and JAM22 counterparts.
Nevertheless, the shapes are very similar.
Finally, we notice that the NNPDFpol1.1 and JAM22 find a polarized gluon which is significantly larger than zero
at medium/large $x$, as confirmed also by the DSSV analysis~\cite{deFlorian:2014yva},
while in our parametrization $\Delta g$ is assumed to be zero at the fit scale $Q^2=4$GeV$^2$.

All in all, the agreement found is rather remarkable,
given that our set is constructed without any information from polarized data.
While as already mentioned performing a simultaneous fit of unpolarized and polarized data
requires a significant amount of work, we can try to include some information on polarized PDFs
in our fit with a little effort.
Specifically, isospin symmetry (which is expected to hold to high accuracy)
implies the so-called triplet sum rule
\beq
\int_0^1 dx\, \Delta T_3(x,\mu^2) = \abs{\frac{g_A}{g_V}} = 1.2754 \pm 0.0013
\eeq
where $g_A$ and $g_V$ are the axial and vector electroweak couplings
which can be derived from neutron decay.
The reported experimental value is taken from the latest PDG average~\cite{ParticleDataGroup:2022pth}.
Notably, the first moment on the left-hand side is scale independent, and so the sum rule is valid at any scale.

We have thus performed an additional NNLO fit to HERA data with the QSPDFflex parametrization
in which we impose the triplet sum rule.
Practically, we do so as if it were a datapoint, to account for the experimental uncertainty.
The result of the fit has a slightly worse $\chi^2/$d.o.f.: from 1334/1109 without the sum rule to 1342/1110 with the sum rule.
This deterioration is driven by the charged-current positron subset,
whose partial $\chi^2$ increases from 55 to 62, over 39 datapoints,
while all other subset are essentially unchanged.
Interestingly, the partial $\chi^2$ from the triplet sum rule is extremely small (close to zero),
despite the very small (permille) uncertainty on the experimental value,
showing that the QSPDFflex parametrization is able to easily accommodate such a physical constraint.

The resulting polarized PDFs are shown in the same figure~\ref{fig:pol}, in red.
The simple imposition of the sum rule improves the agreement with NNPDFpol1.1 and JAM22 significantly.
Among the quarks, only the shape of $\Delta u$, which in turn affects $\Delta T_3$,
is not compatible with NNPDFpol1.1, although it is very close.
We are thus tempted to hope that if a single constraint was able to achieve such a good agreement,
the inclusion of polarized data in the fit
could further improve the agreement without a significant deterioration of the fit quality.
We must also note that the polarized PDF determinations of NNPDFpol1.1, JAM22 and DSSV
have been obtained using NLO theory,
while our fit is NNLO accurate: this difference may also contribute to the disagreement.
We also stress that the unpolarized PDFs of our new fit
are essentially unchanged with respect to the QSPDFflex set without the triplet sum rule.

The difference on the polarized gluon PDF deserves a separate discussion.
The assumption of a zero gluon polarization at the fit scale $\mu_0$
descends from the equilibrium condition which is at the core of the statistical model,
which in turn implies a vanishing potential for the gluon PDF leading to the Plank distribution Eq.~\eqref{eq:gluonPDF}.
However, DGLAP evolution brings partons away from equilibrium,
as it only describes splittings and not recombination.
Indeed, recombination is supposed to be a necessary ingredient only
in a strongly interacting regime, which may happen either at very high density
(i.e.\ at very small $x$, leading to saturation) or in the non-perturbative region at low $Q^2$.
Therefore, the QSPDF(flex) parametrization is best suited to describe PDFs at the border
between non-perturbative and perturbative regimes, namely
at the border between strong dynamics and DGLAP realm.
The choice $\mu_0=2$~GeV for this border may not be optimal ---
indeed, admittedly $2$~GeV is a scale which is certainly in the perturbative regime.
This choice was used in the original literature~\cite{Bourrely:2001du,Buccella:2019yij}
and justified a posteriori by the ability to obtain a good description of the data.
However, meanwhile data have improved, leading in particular to the striking evidence of a polarized gluon
at 2~GeV from RICH data, which may be due to DGLAP evolution from a lower equilibrium scale.

\begin{figure}[t]
  \centering
  \includegraphics[width=0.328\textwidth,page=1]{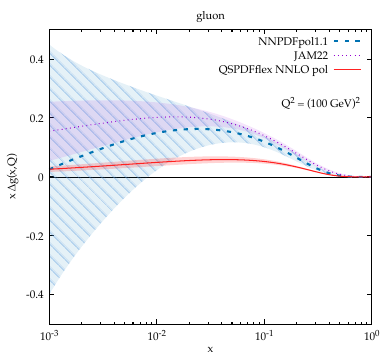}
  \includegraphics[width=0.328\textwidth,page=7]{PDF_comparison__pol_evol.pdf}
  \includegraphics[width=0.328\textwidth,page=8]{PDF_comparison__pol_evol.pdf}
  \\
  \includegraphics[width=0.328\textwidth,page=2]{PDF_comparison__pol_evol.pdf}
  \includegraphics[width=0.328\textwidth,page=5]{PDF_comparison__pol_evol.pdf}
  \includegraphics[width=0.328\textwidth,page=6]{PDF_comparison__pol_evol.pdf}
  \caption{Same as figure~\ref{fig:pol}, but at $Q=100$~GeV and showing only the QSPDFflex NNLO result
    with the triplet sum rule compared with the other public sets.}
  \label{fig:pol100}
\end{figure}

To understand if this may be the case, we have tried to evolve the PDFs to a higher scale,
specifically to $Q=100$~GeV.
For our set (we now consider only the one obtained with the triplet sum rule)
we used \texttt{APFEL++}~\cite{Bertone:2013vaa,Bertone:2017gds}
to perform the polarized evolution (at NLO).
The resulting PDFs are shown in figure~\ref{fig:pol100}.
We notice that the QSPDFflex polarized gluon, thanks to the evolution, is no longer vanishing,
and in particular it is positive as the NNPDFpol1.1 and JAM22 ones.
It is still smaller than those and not compatible with them at medium $x$,
as a consequence of the differences at the starting scale.
We also observe that the agreement of quark PDFs improves after evolution.

We thus conclude that it is very likely that
the optimal scale for the QSPDF(flex) parametrization is a smaller one.
This has been explored in the literature on the statistical parametrization.
For example, Refs.~\cite{Bourrely:2014uha,Bourrely:2015kla} use $\mu_0=1$~GeV.
Nevertheless, they find that even starting from such a low scale data are better described
if a non-zero gluon polarization is assumed at the fit scale.
Moreover, DGLAP evolution from 1~GeV to 2~GeV at fixed order, be it NLO or NNLO or even higher,
is likely inaccurate due to the large value of $\as$ in this region that makes missing higher order
contributions very sizable.
Choosing a starting scale below 1~GeV would make this issue even worse and must thus be avoided.
We conclude that further studies are needed to understand what the best strategy could be.
Perhaps allowing a non-zero gluon polarization is the simplest solution, but a functional
form compatible with the model assumptions must be worked out.

\subsection{Future directions}

The results presented so far show that the QSPDF(flex) parametrization
has a number of virtues, due to the possibility of producing
unpolarized and polarized PDFs with sensible physical behaviour
with a very small number of parameters.
For this reason, on top of being a useful complement to standard parametrizations,
it is worth considering it for further studies.
In this respect, it would be interesting to perform a global fit,
including unpolarized data from other DIS experiment as well as collider data from
Tevatron and LHC, and eventually also polarized data.

As we already mentioned, for the moment this task cannot be performed straight away in \xfitter\
due to the limited amount of available datasets
and the lack of theoretical predictions for polarized observables.
In any case, it is easy to foresee that the simple parametrizations
that we have considered so far will not be suitable to describe a large variety of data.
For such a study, the statistical model parametrization must be made more flexible.
We can increase the flexibility of the parametrization in three ways:
\begin{itemize}
\item adopting Eq.~\eqref{eq:CudY} rather than Eq.~\eqref{eq:CudX}
  for defining the coefficients $C^\uda, \bar C^\uda$,
  which introduces two extra parameters for each quark flavour to be fitted
  and thus gives more flexibility to model the medium/high-$x$ region of quark PDFs;\footnote
  {In fact, Eq.~\eqref{eq:CudX} is an approximation of Eq.~\eqref{eq:CudY}, so the use of the latter
    is the most consistent way of using the model to parametrize quark PDFs.}

\item modifying the low-$x$ behaviour of quark PDFs in the same way we did for the gluon,
  Eq.~\eqref{eq:gluonPDFflex}, namely multiplying each term in the parametrization by
  a polynomial in $\log x$, which gives much more flexibility at small $x$
    without altering the large-$x$ region where the model is physically motivated;

\item providing an independent parametrization for the strange quark PDF,
  needed for fitting data beyond HERA,
  using e.g.\ the parametrization proposed in Ref.~\cite{Bourrely:2007if}
  in the context of the statistical model;

\item changing (lowering) the fit scale $\mu_0$ at which PDFs are parametrize
    to find an optimal value where the assumptions of the model are reasonably satisfied,
    and possibly introduce a parametrization for polarized gluons (see discussion at the end of section~\ref{sec:pol}).

\end{itemize}
All these modifications are perfectly consistent with the model hypotheses,
and can thus be viewed as a natural extension of the study in this work.
As a result of this extension, the parametrization would depend on many more parameters,
reducing some of the advantages of the statistical parametrization, but keeping
the important physical properties, e.g.\ the possibility of determining unpolarized and polarized PDFs
with the same parameters.
Note also that some of the new parameters (in particular those modelling the small-$x$ behaviour)
may be redundant and could possibly be eliminated,
but this can only be decided after performing the fit.

The extended parametrization proposed above would be much more flexible,
making it comparable with other fixed-form flexible parametrizations on the market.
Despite the flexibility, such an improved QSPDF parametrization would still
differ from other parametrizations in one key aspect: the $x\to1$ behaviour.
Indeed, the statistical model is characterized by a non-vanishing limit in $x=1$, Eq.~\eqref{eq:QSx=1},
while all other parametrizations, including the very flexible NNPDF one,
assume a vanishing power-like behaviour of the form $(1-x)^\alpha$ with $\alpha>0$.
We have already seen in section~\ref{sec:flex} that differences are significant
for $x\gtrsim0.7$: the inclusion of data sensitive to such high values of $x$,
such as LHC jet data or future EIC data,
would thus allow us to test which functional form is more suitable for their description.
On top of looking at the fit quality, it could be possible to perform a simple test
to verify whether the data prefer vanishing or non-vanishing PDFs in $x=1$:
after having fitted high-$x$ data with the extended QSPDF parametrization,
one could multiply each PDF parametrization by a $(1-x)^\alpha$ factor,
with the same value of $\alpha$ for all PDFs,\footnote
{Using different values of $\alpha$ for each PDFs may lead to overfitting, as the extended QSPDF parametrization
  is already rather flexible at high $x$.
  Moreover here the goal is to understand whether data prefer vanishing or a non-vanishing PDFs
  in $x=1$, and this is better tested by the simplest modification of the parametrization
  that adds the least flexibility.}
and fit data again.
If the fit selects a (positive) value of $\alpha$ significantly different from zero
\emph{and} the $\chi^2$ reduces by more than one unity (corresponding to the extra fitted parameter),
then one must conclude that data prefer vanishing PDFs in $x=1$.
Conversely, if at least one of the conditions above are not satisfied, the non-vanishing
behaviour predicted by the statistical model has to be considered as compatible with the data.

On top of this high-$x$ test, another powerful validation of the model consists
in the ability of fitting in a satisfactory way both unpolarized and polarized data.
The results of section~\ref{sec:pol} are very encouraging, but the actual test
can only come once polarized and more unpolarized data are included in the fit.
As discussed in section~\ref{sec:pol}, a key issue to be faced is the polarized gluon PDF:
while at equilibrium it makes sense to assume zero polarization, it is not obvious that
it is possible to really start the fit at the equilibrium scale, thus requiring the introduction
of a parametrization for the polarized gluon PDF.
Finding such a parametrization in a way that is compatible with the statistical model
requires a theory study that is left to a future work.

\section{Conclusions}
\label{sec:conclusions}

In this work we have considered a PDF parametrization inspired by a statistical model of the proton dynamics
and tested it in fits to HERA data through the public \xfitter\ code.
The idea behind the use of such parametrization is the opposite of the standard practice:
while usually one tries to consider very flexible functional forms to reduce the parametrization bias,
here we consider very biased functional forms to reduce the number of fitting parameters.
The hope is that the bias introduced in the parametrization be justified by the physical
model which the functional form is derived from, thus leading to a reasonably good fit
with all the advantages of a small set of parameters.
To our knowledge, this is the first PDF fit based on the statistical model performed at NNLO and NNLO+NLL$x$
accuracy (previous ones were only NLO accurate).

We have considered two versions of the parametrization.
One is the simplest original version coming from the model~\cite{Bourrely:2001du},
denoted QSPDF, which has 9 free parameters.
The other one is a variant in which we add more flexibility to the gluon at small $x$,
denoted QSPDFflex, which depends on 11 free parameters.
The QSPDF parametrization allows to reasonably describe HERA data,
but the quality of the fit is not particularly high, due to an extremely limited functional form
of the gluon PDF which does not allow to describe low-$x$ data well.
The QSPDFflex parametrization, instead, leads to a good description of the data
comparable with other PDF parametrizations which depend on more parameters.
For instance, the $\chi^2$ obtained with QSPDFflex is essentially the same as that obtained
with a HERAPDF-like parametrization.
We can thus conclude that the QSPDFflex parametrization is working well
and it can provide a useful complement to standard parametrization,
e.g.\ to study parametrization bias
(or equivalently to estimate a parametrization uncertainty)
or as a new starting point for constructing more flexible parametrizations.

In terms of PDF comparison, we have noticed that the shapes predicted by QSPDF and QSPDFflex differ
in several respects from the ones of a HERAPDF-like parametrization.
The valence distributions differ noticeably, and the QSPDF(flex) results are in better agreement
with other PDF sets (NNPDF, BG) based on more flexible parametrizations.
Conversely, at small $x$ the QSPDF shapes, in particular for the gluon,
are very constrained and thus differ from all other PDF sets on the market.
The QSPDFflex set cures this problem, allowing the gluon to behave in better agreement with other public sets.
It has to be noticed however that the shape of the gluon at small $x$ is very sensitive also
to the dataset and the theory ingredients, and so different sets may differ visibly.

Related to this last observation, we have also considered fits in which theoretical predictions
are supplemented by small-$x$ resummation.
In these fits we expected the gluon to be more in agreement among different sets,
because the inclusion of the resummation stabilises the fit in the small-$x$ region
and predicts a gluon that tends to rise more steeply.
We have indeed found a good agreement between various resummed fits on the market
and our QSPDF(flex) fit, as well as a HERAPDF-like fit. The inclusion of small-$x$ resummation
also produces a reduction of the $\chi^2$,
in agreement with previous studies~\cite{Ball:2017otu,xFitterDevelopersTeam:2018hym,Bonvini:2019wxf}.

Finally, we have made some considerations on how our results impact
the statistical model itself.
The success of the QSPDF(flex) parametrization in describing HERA data can be seen as a sort of validation of the model.
Also the fact that QSPDFflex predicts a positive $\bar d-\bar u$ distribution
even if the HERA data do not contain enough information to separate these flavours
can be seen as an indication that the model is reliable.
We have verified that this prediction is stable upon inclusion of additional data at high $x$.
Moreover, if we were to trust the model, then it would allow to describe
with the very same parameters polarized PDFs as well.
We have verified that our fit to unpolarized data
produces polarized PDFs that are remarkably similar to those fitted from data,
and the simple inclusion of the triplet sum rule in the fit further improves the agreement.
This shows that it is probably possible to simultaneously fit with the same parameters
both unpolarized and polarized data, and we leave this investigation to future work.

The QSPDF and QSPDFflex sets at NNLO and NNLO plus small-$x$ resummation can be downloaded at the address
\href{http://l.infn.it/qspdf}{\texttt{l.infn.it/qspdf}}.
These sets should not be regarded as general-purpose PDFs,
because they have been obtained using a reduced dataset and the parametrization adopted is very minimal.
When trying to fit more data, including other DIS experiments and collider (LHC) measurements,
it is very likely that the very simple parametrizations proposed, even the QSPDFflex one,
will not be able to achieve a satisfactory fit quality.
To get a competitive fit, it is possible to improve the parametrization in three respects.
First, it is possible to parametrize the strange PDF independently, using e.g.\ the formulation of Ref.~\cite{Bourrely:2007if}.
Second, we can give more flexibility to the high-$x$ region by
using less constrained values for the $C^\uda_q$ parameters,
e.g.\ introducing the so-called ``transverse potentials'', Eq.~\eqref{eq:CudY}.
Third, we can further model the small-$x$ region by introducing polynomials in $\log x$~\cite{Bonvini:2019wxf}
as we did for the gluon, Eq.~\eqref{eq:gluonPDFflex}, also for the quark PDFs.
Including also polarized data in the fit, such as those from RHIC,
further requires an extension of the model to introduce a polarized gluon PDF.
We leave the study of these extensions to future work.

\acknowledgments{We are grateful to Francesco Tramontano for collaboration during various phases of this work.
  We thank Mandy Cooper-Sarkar and Sasha Glazov for useful comments on the manuscript.
  The work of MB is partly supported by the Italian Ministry of University and Research (MUR) grant PRIN 2022SNA23K
  funded by the European Union -- Next Generation EU.}

\appendix

\section{Implementation of the sum rules}
\label{sec:sumrules}

The parameters of the fit are constrained by the quark number and momentum sum rules:
\begin{align}
    2&=\int_0^1 dx\, u_v(x,\mu_0^2), \label{eq:sumrule1}\\
    1&=\int_0^1 dx\, d_v(x,\mu_0^2), \label{eq:sumrule2}\\
    1&=\int_0^1 dx\, x[g(x,\mu_0^2)+u(x,\mu_0^2)+\bar u(x,\mu_0^2) +d(x,\mu_0^2)+\bar d(x,\mu_0^2)+s(x,\mu_0^2)+\bar s(x,\mu_0^2)].
\end{align}
Usually, they are used to fix the normalization of $u_v$, $d_v$ and the gluon respectively.
However, the parametrization Eq.~\eqref{eq:PDFpar} does not have an overall normalization factor
for valence distributions, making the implementation of the quark number sum rules non trivial.

The parametrization Eq.~\eqref{eq:PDFpar} for valence quarks has the form of the sum of two contributions,
one proportional to the parameter $A$ and the other to $\bar A$.
Crucially, these two parameters are the same for the up and the down quarks.
Therefore, it is still possible to use the quark number sum rules Eqs.~\eqref{eq:sumrule1} and \eqref{eq:sumrule2}
to determine $A$ and $\bar A$, but in order to do so we have to solve the algebraic system
\beq
\mqty(2\\1) = \mqty(K_u & \Kb_u\\ K_d & \Kb_d)\mqty(A \\ \Ab) \equiv \mathcal{K}\mqty(A\\ \Ab)
\eeq
with
\begin{align}
K_q &= \int_0^1 \frac{\dd{x}}x\, \[C^\uparr_q h_+(x;b, X^\uparr_q ) + C^\dwarr_q h_+(x;b, X^\dwarr_q )\] ,\\ 
\Kb_q &= -\int_0^1 \frac{\dd{x}}x\, \[\Cb^\uparr_q h_+(x;\bb, -X^\dwarr_q ) + \Cb^\dwarr_q h_+(x;\bb, -X^\uparr_q )\] .
\end{align}
The solution is given by
\begin{equation}\label{eq:AAbar}
  A = \frac{1}{\det(\mathcal{K})} \det(\mqty{2 & \Kb_u \\ 1 & \Kb_d}) \, \qquad
  \Ab = \frac{1}{\det(\mathcal{K})} \det(\mqty{K_u & 2 \\ K_d & 1}) \, .
\end{equation}
In order to implement this procedure in \xfitter, we have introduced a new PDF decomposition
in which the valence parts of quarks and antiquarks (those proportional to $A$ and $\bar A$ respectively)
are considered as independent PDFs, as well as the diffractive (sea) term.
After the determination of $A$ and $\bar A$ according to Eq.~\eqref{eq:AAbar},
they are combined together to form the parametrization for valence quarks and antiquarks as
in Eq.~\eqref{eq:PDFpar}.

The sum rules integrals are computed numerically in \xfitter.
In particular, the integral is divided into two regions,
one in $0.1<x<1$ which is sampled linearly,
and one in $10^{-6}<x<0.1$ which is sampled logarithmically.
Therefore, the generic sum rule integrals are approximated as
\beq\label{eq:sumrulint}
\int_0^1 dx\, x^N f(x,\mu_0^2) \simeq  \int_{x_0}^1 dx\, x^N f(x,\mu_0^2)\qquad (N=0,1)
\eeq
with $x_0=10^{-6}$.
The neglected region below this value is usually harmless.
However, if the integrand is divergent and not integrable in $x=0$,
the approximation still gives a finite value thanks to the cutoff $x_0$.
This may be problematic, as the non-integrability is a manifestation of
bad values of the parameters of the PDFs, which must be avoided in the fit.
The divergence of the integral is thus a way to put barriers to some parameters.

To this end, it is important to improve the approximation Eq.~\eqref{eq:sumrulint}
by adding the contribution below $x_0$.
We assume that all PDFs at small $x$ have a power like behaviour
\beq
f(x,\mu_0^2)\overset{x\to0}\sim \alpha\, x^\beta,
\eeq
which is the case for most parametrizations, including the ones we use in this work.
The integral from 0 to $x_0=10^{-6}$ can then be approximated by
$$
\int_0^{x_0}dx\, x^N f(x,\mu_0^2) \simeq \alpha\,\frac{x_0^{\beta+N+1}}{\beta+N+1}.
$$
We have implemented this additional contribution in \xfitter,
with a numerical extrapolation of $\beta$ and $\alpha$.
In this way, values of $\beta\leq-1-N$ become also practically forbidden,
thus providing the aforementioned barriers to the proper parameters.
Effectively, for the parametrization of Sect.~\ref{sec:model},
this constraint forces the following conditions on the $b$ parameters:
$b,\bb,b_g>0$ and $\tilde b>-1$.

\phantomsection
\addcontentsline{toc}{section}{References}

\bibliographystyle{jhep}
\bibliography{xfitter-QSpdfs-fit}

\providecommand{\href}[2]{#2}\begingroup\raggedright\begin{thebibliography}{100}

\bibitem{H1:2015ubc}
{\scshape H1, ZEUS} collaboration, H.~Abramowicz et~al., \emph{{Combination of
  measurements of inclusive deep inelastic ${e^{\pm }p}$ scattering cross
  sections and QCD analysis of HERA data}},
  \href{https://doi.org/10.1140/epjc/s10052-015-3710-4}{\emph{Eur. Phys. J. C}
  {\bfseries 75} (2015) 580}
  [\href{https://arxiv.org/abs/1506.06042}{{\ttfamily 1506.06042}}].

\bibitem{Alekhin:2017kpj}
S.~Alekhin, J.~Bl\"umlein, S.~Moch and R.~Placakyte, \emph{{Parton distribution
  functions, $\alpha_s$, and heavy-quark masses for LHC Run II}},
  \href{https://doi.org/10.1103/PhysRevD.96.014011}{\emph{Phys. Rev. D}
  {\bfseries 96} (2017) 014011}
  [\href{https://arxiv.org/abs/1701.05838}{{\ttfamily 1701.05838}}].

\bibitem{Hou:2019efy}
T.-J. Hou et~al., \emph{{New CTEQ global analysis of quantum chromodynamics
  with high-precision data from the LHC}},
  \href{https://doi.org/10.1103/PhysRevD.103.014013}{\emph{Phys. Rev. D}
  {\bfseries 103} (2021) 014013}
  [\href{https://arxiv.org/abs/1912.10053}{{\ttfamily 1912.10053}}].

\bibitem{Bailey:2020ooq}
S.~Bailey, T.~Cridge, L.~A. Harland-Lang, A.~D. Martin and R.~S. Thorne,
  \emph{{Parton distributions from LHC, HERA, Tevatron and fixed target data:
  MSHT20 PDFs}},
  \href{https://doi.org/10.1140/epjc/s10052-021-09057-0}{\emph{Eur. Phys. J. C}
  {\bfseries 81} (2021) 341}
  [\href{https://arxiv.org/abs/2012.04684}{{\ttfamily 2012.04684}}].

\bibitem{NNPDF:2021njg}
{\scshape NNPDF} collaboration, R.~D. Ball et~al., \emph{{The path to proton
  structure at 1\% accuracy}},
  \href{https://doi.org/10.1140/epjc/s10052-022-10328-7}{\emph{Eur. Phys. J. C}
  {\bfseries 82} (2022) 428}
  [\href{https://arxiv.org/abs/2109.02653}{{\ttfamily 2109.02653}}].

\bibitem{Accardi:2021ysh}
A.~Accardi, T.~J. Hobbs, X.~Jing and P.~M. Nadolsky, \emph{{Deuterium
  scattering experiments in CTEQ global QCD analyses: a comparative
  investigation}},
  \href{https://doi.org/10.1140/epjc/s10052-021-09318-y}{\emph{Eur. Phys. J. C}
  {\bfseries 81} (2021) 603}
  [\href{https://arxiv.org/abs/2102.01107}{{\ttfamily 2102.01107}}].

\bibitem{ATLAS:2021vod}
{\scshape ATLAS} collaboration, G.~Aad et~al., \emph{{Determination of the
  parton distribution functions of the proton using diverse ATLAS data from
  $pp$ collisions at $\sqrt{s} = 7$, 8 and 13~TeV}},
  \href{https://doi.org/10.1140/epjc/s10052-022-10217-z}{\emph{Eur. Phys. J. C}
  {\bfseries 82} (2022) 438}
  [\href{https://arxiv.org/abs/2112.11266}{{\ttfamily 2112.11266}}].

\bibitem{CMS:2021yzl}
{\scshape CMS} collaboration, A.~Tumasyan et~al., \emph{{Measurement and QCD
  analysis of double-differential inclusive jet cross sections in proton-proton
  collisions at $ \sqrt{s} $ = 13 TeV}},
  \href{https://doi.org/10.1007/JHEP02(2022)142}{\emph{JHEP} {\bfseries 02}
  (2022) 142} [\href{https://arxiv.org/abs/2111.10431}{{\ttfamily
  2111.10431}}].

\bibitem{Ji:2013dva}
X.~Ji, \emph{{Parton Physics on a Euclidean Lattice}},
  \href{https://doi.org/10.1103/PhysRevLett.110.262002}{\emph{Phys. Rev. Lett.}
  {\bfseries 110} (2013) 262002}
  [\href{https://arxiv.org/abs/1305.1539}{{\ttfamily 1305.1539}}].

\bibitem{Radyushkin:2017cyf}
A.~V. Radyushkin, \emph{{Quasi-parton distribution functions, momentum
  distributions, and pseudo-parton distribution functions}},
  \href{https://doi.org/10.1103/PhysRevD.96.034025}{\emph{Phys. Rev. D}
  {\bfseries 96} (2017) 034025}
  [\href{https://arxiv.org/abs/1705.01488}{{\ttfamily 1705.01488}}].

\bibitem{Lin:2017snn}
H.-W. Lin et~al., \emph{{Parton distributions and lattice QCD calculations: a
  community white paper}},
  \href{https://doi.org/10.1016/j.ppnp.2018.01.007}{\emph{Prog. Part. Nucl.
  Phys.} {\bfseries 100} (2018) 107}
  [\href{https://arxiv.org/abs/1711.07916}{{\ttfamily 1711.07916}}].

\bibitem{Altarelli:1977zs}
G.~Altarelli and G.~Parisi, \emph{{Asymptotic Freedom in Parton Language}},
  \href{https://doi.org/10.1016/0550-3213(77)90384-4}{\emph{Nucl. Phys. B}
  {\bfseries 126} (1977) 298}.

\bibitem{Gribov:1972ri}
V.~N. Gribov and L.~N. Lipatov, \emph{{Deep inelastic e p scattering in
  perturbation theory}}, {\emph{Sov. J. Nucl. Phys.} {\bfseries 15} (1972)
  438}.

\bibitem{Dokshitzer:1977sg}
Y.~L. Dokshitzer, \emph{{Calculation of the Structure Functions for Deep
  Inelastic Scattering and e+ e- Annihilation by Perturbation Theory in Quantum
  Chromodynamics.}}, {\emph{Sov. Phys. JETP} {\bfseries 46} (1977) 641}.

\bibitem{NNPDF:2017mvq}
{\scshape NNPDF} collaboration, R.~D. Ball et~al., \emph{{Parton distributions
  from high-precision collider data}},
  \href{https://doi.org/10.1140/epjc/s10052-017-5199-5}{\emph{Eur. Phys. J. C}
  {\bfseries 77} (2017) 663}
  [\href{https://arxiv.org/abs/1706.00428}{{\ttfamily 1706.00428}}].

\bibitem{Bourrely:1993wq}
C.~Bourrely, F.~Buccella, G.~Miele, G, G.~Migliore, J.~Soffer et~al.,
  \emph{{Fermi-Dirac distributions for quark partons}},
  \href{https://doi.org/10.1007/BF01555903}{\emph{Z. Phys. C} {\bfseries 62}
  (1994) 431} [\href{https://arxiv.org/abs/hep-ph/9410375}{{\ttfamily
  hep-ph/9410375}}].

\bibitem{Buccella:1996kb}
F.~Buccella, G.~Miele and N.~Tancredi, \emph{{Quantum statistical parton
  distributions and the spin crisis}},
  \href{https://doi.org/10.1143/PTP.96.749}{\emph{Prog. Theor. Phys.}
  {\bfseries 96} (1996) 749}
  [\href{https://arxiv.org/abs/hep-ph/9604230}{{\ttfamily hep-ph/9604230}}].

\bibitem{Bourrely:2001du}
C.~Bourrely, J.~Soffer and F.~Buccella, \emph{{A Statistical approach for
  polarized parton distributions}},
  \href{https://doi.org/10.1007/s100520100855}{\emph{Eur. Phys. J. C}
  {\bfseries 23} (2002) 487}
  [\href{https://arxiv.org/abs/hep-ph/0109160}{{\ttfamily hep-ph/0109160}}].

\bibitem{Bourrely:2002xm}
C.~Bourrely, F.~Buccella and J.~Soffer, \emph{{Recent tests for the statistical
  parton distributions}},
  \href{https://doi.org/10.1142/S0217732303009861}{\emph{Mod. Phys. Lett. A}
  {\bfseries 18} (2003) 771}
  [\href{https://arxiv.org/abs/hep-ph/0211389}{{\ttfamily hep-ph/0211389}}].

\bibitem{Bourrely:2005kw}
C.~R.~V. Bourrely, J.~Soffer and F.~Buccella, \emph{{The Statistical parton
  distributions: Status and prospects}},
  \href{https://doi.org/10.1140/epjc/s2005-02205-2}{\emph{Eur. Phys. J. C}
  {\bfseries 41} (2005) 327}
  [\href{https://arxiv.org/abs/hep-ph/0502180}{{\ttfamily hep-ph/0502180}}].

\bibitem{Bourrely:2005tp}
C.~Bourrely, J.~Soffer and F.~Buccella, \emph{{The Extension to the transverse
  momentum of the statistical parton distributions}},
  \href{https://doi.org/10.1142/S0217732306019244}{\emph{Mod. Phys. Lett. A}
  {\bfseries 21} (2006) 143}
  [\href{https://arxiv.org/abs/hep-ph/0507328}{{\ttfamily hep-ph/0507328}}].

\bibitem{Bourrely:2010ng}
C.~Bourrely, F.~Buccella and J.~Soffer, \emph{{Semiinclusive DIS cross sections
  and spin asymmetries in the quantum statistical parton distributions
  approach}}, \href{https://doi.org/10.1103/PhysRevD.83.074008}{\emph{Phys.
  Rev. D} {\bfseries 83} (2011) 074008}
  [\href{https://arxiv.org/abs/1008.5322}{{\ttfamily 1008.5322}}].

\bibitem{Bourrely:2013yti}
C.~Bourrely, F.~Buccella and J.~Soffer, \emph{{The transverse momentum
  dependent statistical parton distributions revisited}},
  \href{https://doi.org/10.1142/S0217751X13500267}{\emph{Int. J. Mod. Phys. A}
  {\bfseries 28} (2013) 1350026}
  [\href{https://arxiv.org/abs/1302.4281}{{\ttfamily 1302.4281}}].

\bibitem{Bourrely:2013qfa}
C.~Bourrely, F.~Buccella and J.~Soffer, \emph{{$W^{\pm}$ bosons production in
  the quantum statistical parton distributions approach}},
  \href{https://doi.org/10.1016/j.physletb.2013.08.045}{\emph{Phys. Lett. B}
  {\bfseries 726} (2013) 296}
  [\href{https://arxiv.org/abs/1308.3567}{{\ttfamily 1308.3567}}].

\bibitem{Bourrely:2014uha}
C.~Bourrely and J.~Soffer, \emph{{Statistical description of the proton spin
  with a large gluon helicity distribution}},
  \href{https://doi.org/10.1016/j.physletb.2014.11.044}{\emph{Phys. Lett. B}
  {\bfseries 740} (2015) 168}
  [\href{https://arxiv.org/abs/1408.7057}{{\ttfamily 1408.7057}}].

\bibitem{Buccella:2014wpa}
F.~Buccella and S.~Sohaily, \emph{{A Check-up for the Statistical Parton
  Model}}, \href{https://doi.org/10.1142/S021773231550203X}{\emph{Mod. Phys.
  Lett. A} {\bfseries 30} (2015) 1550203}
  [\href{https://arxiv.org/abs/1412.7683}{{\ttfamily 1412.7683}}].

\bibitem{Bourrely:2015kla}
C.~Bourrely and J.~Soffer, \emph{{New developments in the statistical approach
  of parton distributions: tests and predictions up to LHC energies}},
  \href{https://doi.org/10.1016/j.nuclphysa.2015.06.018}{\emph{Nucl. Phys. A}
  {\bfseries 941} (2015) 307}
  [\href{https://arxiv.org/abs/1502.02517}{{\ttfamily 1502.02517}}].

\bibitem{Bourrely:2018yck}
C.~Bourrely and J.~Soffer, \emph{{Statistical approach of pion parton
  distributions from Drell\textendash{}Yan process}},
  \href{https://doi.org/10.1016/j.nuclphysa.2018.07.003}{\emph{Nucl. Phys. A}
  {\bfseries 981} (2019) 118}
  [\href{https://arxiv.org/abs/1802.03153}{{\ttfamily 1802.03153}}].

\bibitem{Soffer:2019gbb}
J.~Soffer and C.~Bourrely, \emph{{On the flavor structure of the light-quark
  sea distributions}},
  \href{https://doi.org/10.1016/j.nuclphysa.2019.08.001}{\emph{Nucl. Phys. A}
  {\bfseries 991} (2019) 121607}.

\bibitem{Buccella:2019yij}
F.~Buccella, S.~Sohaily and F.~Tramontano, \emph{{Low $Q^2$ boundary conditions
  for DGLAP equations dictated by quantum statistical mechanics}},
  \href{https://doi.org/10.1088/1742-5468/ab054e}{\emph{J. Stat. Mech.}
  {\bfseries 1907} (2019) 073302}.

\bibitem{Bourrely:2020izp}
C.~Bourrely, F.~Buccella and J.-C. Peng, \emph{{A new extraction of pion parton
  distributions in the statistical model}},
  \href{https://doi.org/10.1016/j.physletb.2020.136021}{\emph{Phys. Lett. B}
  {\bfseries 813} (2021) 136021}
  [\href{https://arxiv.org/abs/2008.05703}{{\ttfamily 2008.05703}}].

\bibitem{Bourrely:2022mjf}
C.~Bourrely, W.-C. Chang and J.-C. Peng, \emph{{Pion Partonic Distributions in
  the Statistical Model from Pion-induced Drell-Yan and $J/\Psi$ Production
  Data}}, \href{https://doi.org/10.1103/PhysRevD.105.076018}{\emph{Phys. Rev.
  D} {\bfseries 105} (2022) 076018}
  [\href{https://arxiv.org/abs/2202.12547}{{\ttfamily 2202.12547}}].

\bibitem{Buccella:2022tmb}
F.~Buccella, \emph{{Status of the Quantum Statistical Approach to the Parton
  Distributions}}, \href{https://doi.org/10.22323/1.406.0003}{\emph{PoS}
  {\bfseries CORFU2021} (2022) 003}.

\bibitem{Bellantuono:2022hqp}
L.~Bellantuono, R.~Bellotti and F.~Buccella, \emph{{Planck formula for the
  gluon parton distribution in the proton}},
  \href{https://doi.org/10.1142/S0217732323500396}{\emph{Mod. Phys. Lett. A}
  {\bfseries 38} (2023) 2350039}
  [\href{https://arxiv.org/abs/2201.07640}{{\ttfamily 2201.07640}}].

\bibitem{Bourrely:2023yzi}
C.~Bourrely, F.~Buccella, W.-C. Chang and J.-C. Peng, \emph{{Extraction of Kaon
  Partonic Distribution Functions from Drell-Yan and $J/\psi$ Production
  Data}},  [\href{https://arxiv.org/abs/2305.18117}{{\ttfamily 2305.18117}}].

\bibitem{Silvetti:2024fdi}
F.~Silvetti, \emph{{Resummation phenomenology and PDF determination for
  precision QCD at the LHC}}, phd thesis, Sapienza, Universitiy of Rome, 9,
  2023.
\newblock \href{https://arxiv.org/abs/2403.20315}{{\ttfamily 2403.20315}}.

\bibitem{deFlorian:2014yva}
D.~de~Florian, R.~Sassot, M.~Stratmann and W.~Vogelsang, \emph{{Evidence for
  polarization of gluons in the proton}},
  \href{https://doi.org/10.1103/PhysRevLett.113.012001}{\emph{Phys. Rev. Lett.}
  {\bfseries 113} (2014) 012001}
  [\href{https://arxiv.org/abs/1404.4293}{{\ttfamily 1404.4293}}].

\bibitem{xFitterDevelopersTeam:2017fzy}
{\scshape xFitter Developers Team} collaboration, V.~Bertone et~al.,
  \emph{{Impact of the heavy quark matching scales in PDF fits}},
  \href{https://doi.org/10.1140/epjc/s10052-017-5407-3}{\emph{Eur. Phys. J. C}
  {\bfseries 77} (2017) 837}
  [\href{https://arxiv.org/abs/1707.05343}{{\ttfamily 1707.05343}}].

\bibitem{Bertone:2013vaa}
{\scshape APFEL} collaboration, V.~Bertone, S.~Carrazza and J.~Rojo,
  \emph{{APFEL: A PDF Evolution Library with QED corrections}},
  \href{https://doi.org/10.1016/j.cpc.2014.03.007}{\emph{Comput. Phys. Commun.}
  {\bfseries 185} (2014) 1647}
  [\href{https://arxiv.org/abs/1310.1394}{{\ttfamily 1310.1394}}].

\bibitem{Botje:2010ay}
M.~Botje, \emph{{QCDNUM: Fast QCD Evolution and Convolution}},
  \href{https://doi.org/10.1016/j.cpc.2010.10.020}{\emph{Comput. Phys. Commun.}
  {\bfseries 182} (2011) 490}
  [\href{https://arxiv.org/abs/1005.1481}{{\ttfamily 1005.1481}}].

\bibitem{Thorne:1997ga}
R.~S. Thorne and R.~G. Roberts, \emph{{An Ordered analysis of heavy flavor
  production in deep inelastic scattering}},
  \href{https://doi.org/10.1103/PhysRevD.57.6871}{\emph{Phys. Rev. D}
  {\bfseries 57} (1998) 6871}
  [\href{https://arxiv.org/abs/hep-ph/9709442}{{\ttfamily hep-ph/9709442}}].

\bibitem{Thorne:2006qt}
R.~S. Thorne, \emph{{A Variable-flavor number scheme for NNLO}},
  \href{https://doi.org/10.1103/PhysRevD.73.054019}{\emph{Phys. Rev. D}
  {\bfseries 73} (2006) 054019}
  [\href{https://arxiv.org/abs/hep-ph/0601245}{{\ttfamily hep-ph/0601245}}].

\bibitem{Thorne:2012az}
R.~S. Thorne, \emph{{Effect of changes of variable flavor number scheme on
  parton distribution functions and predicted cross sections}},
  \href{https://doi.org/10.1103/PhysRevD.86.074017}{\emph{Phys. Rev. D}
  {\bfseries 86} (2012) 074017}
  [\href{https://arxiv.org/abs/1201.6180}{{\ttfamily 1201.6180}}].

\bibitem{Forte:2010ta}
S.~Forte, E.~Laenen, P.~Nason and J.~Rojo, \emph{{Heavy quarks in
  deep-inelastic scattering}},
  \href{https://doi.org/10.1016/j.nuclphysb.2010.03.014}{\emph{Nucl. Phys. B}
  {\bfseries 834} (2010) 116}
  [\href{https://arxiv.org/abs/1001.2312}{{\ttfamily 1001.2312}}].

\bibitem{Harland-Lang:2016yfn}
L.~A. Harland-Lang, A.~D. Martin, P.~Motylinski and R.~S. Thorne, \emph{{The
  impact of the final HERA combined data on PDFs obtained from a global fit}},
  \href{https://doi.org/10.1140/epjc/s10052-016-4020-1}{\emph{Eur. Phys. J. C}
  {\bfseries 76} (2016) 186}
  [\href{https://arxiv.org/abs/1601.03413}{{\ttfamily 1601.03413}}].

\bibitem{Abt:2016vjh}
I.~Abt, A.~M. Cooper-Sarkar, B.~Foster, V.~Myronenko, K.~Wichmann and M.~Wing,
  \emph{{Study of HERA ep data at low Q$^2$ and low $x_{Bj}$ and the need for
  higher-twist corrections to standard perturbative QCD fits}},
  \href{https://doi.org/10.1103/PhysRevD.94.034032}{\emph{Phys. Rev. D}
  {\bfseries 94} (2016) 034032}
  [\href{https://arxiv.org/abs/1604.02299}{{\ttfamily 1604.02299}}].

\bibitem{Ball:2017otu}
R.~D. Ball, V.~Bertone, M.~Bonvini, S.~Marzani, J.~Rojo and L.~Rottoli,
  \emph{{Parton distributions with small-x resummation: evidence for BFKL
  dynamics in HERA data}},
  \href{https://doi.org/10.1140/epjc/s10052-018-5774-4}{\emph{Eur. Phys. J. C}
  {\bfseries 78} (2018) 321}
  [\href{https://arxiv.org/abs/1710.05935}{{\ttfamily 1710.05935}}].

\bibitem{xFitterDevelopersTeam:2018hym}
{\scshape xFitter Developers' Team} collaboration, H.~Abdolmaleki et~al.,
  \emph{{Impact of low-$x$ resummation on QCD analysis of HERA data}},
  \href{https://doi.org/10.1140/epjc/s10052-018-6090-8}{\emph{Eur. Phys. J. C}
  {\bfseries 78} (2018) 621}
  [\href{https://arxiv.org/abs/1802.00064}{{\ttfamily 1802.00064}}].

\bibitem{Pumplin:2000vx}
J.~Pumplin, D.~R. Stump and W.~K. Tung, \emph{{Multivariate fitting and the
  error matrix in global analysis of data}},
  \href{https://doi.org/10.1103/PhysRevD.65.014011}{\emph{Phys. Rev. D}
  {\bfseries 65} (2001) 014011}
  [\href{https://arxiv.org/abs/hep-ph/0008191}{{\ttfamily hep-ph/0008191}}].

\bibitem{James:1975dr}
F.~James and M.~Roos, \emph{{Minuit: A System for Function Minimization and
  Analysis of the Parameter Errors and Correlations}},
  \href{https://doi.org/10.1016/0010-4655(75)90039-9}{\emph{Comput. Phys.
  Commun.} {\bfseries 10} (1975) 343}.

\bibitem{Ball:2014uwa}
{\scshape NNPDF} collaboration, R.~D. Ball et~al., \emph{{Parton distributions
  for the LHC Run II}},
  \href{https://doi.org/10.1007/JHEP04(2015)040}{\emph{JHEP} {\bfseries 04}
  (2015) 040} [\href{https://arxiv.org/abs/1410.8849}{{\ttfamily 1410.8849}}].

\bibitem{Salam:1998tj}
G.~P. Salam, \emph{{A Resummation of large subleading corrections at small x}},
  \href{https://doi.org/10.1088/1126-6708/1998/07/019}{\emph{JHEP} {\bfseries
  07} (1998) 019} [\href{https://arxiv.org/abs/hep-ph/9806482}{{\ttfamily
  hep-ph/9806482}}].

\bibitem{Ciafaloni:1999yw}
M.~Ciafaloni, D.~Colferai and G.~P. Salam, \emph{{Renormalization group
  improved small x equation}},
  \href{https://doi.org/10.1103/PhysRevD.60.114036}{\emph{Phys. Rev. D}
  {\bfseries 60} (1999) 114036}
  [\href{https://arxiv.org/abs/hep-ph/9905566}{{\ttfamily hep-ph/9905566}}].

\bibitem{Ciafaloni:2003kd}
M.~Ciafaloni, D.~Colferai, G.~P. Salam and A.~M. Stasto, \emph{{The Gluon
  splitting function at moderately small x}},
  \href{https://doi.org/10.1016/j.physletb.2004.02.054}{\emph{Phys. Lett. B}
  {\bfseries 587} (2004) 87}
  [\href{https://arxiv.org/abs/hep-ph/0311325}{{\ttfamily hep-ph/0311325}}].

\bibitem{Ciafaloni:2003rd}
M.~Ciafaloni, D.~Colferai, G.~P. Salam and A.~M. Stasto, \emph{{Renormalization
  group improved small x Green's function}},
  \href{https://doi.org/10.1103/PhysRevD.68.114003}{\emph{Phys. Rev. D}
  {\bfseries 68} (2003) 114003}
  [\href{https://arxiv.org/abs/hep-ph/0307188}{{\ttfamily hep-ph/0307188}}].

\bibitem{Ciafaloni:2007gf}
M.~Ciafaloni, D.~Colferai, G.~P. Salam and A.~M. Stasto, \emph{{A Matrix
  formulation for small-x singlet evolution}},
  \href{https://doi.org/10.1088/1126-6708/2007/08/046}{\emph{JHEP} {\bfseries
  08} (2007) 046} [\href{https://arxiv.org/abs/0707.1453}{{\ttfamily
  0707.1453}}].

\bibitem{Ball:1995vc}
R.~D. Ball and S.~Forte, \emph{{Summation of leading logarithms at small x}},
  \href{https://doi.org/10.1016/0370-2693(95)00395-2}{\emph{Phys. Lett. B}
  {\bfseries 351} (1995) 313}
  [\href{https://arxiv.org/abs/hep-ph/9501231}{{\ttfamily hep-ph/9501231}}].

\bibitem{Ball:1997vf}
R.~D. Ball and S.~Forte, \emph{{Asymptotically free partons at high-energy}},
  \href{https://doi.org/10.1016/S0370-2693(97)00625-4}{\emph{Phys. Lett. B}
  {\bfseries 405} (1997) 317}
  [\href{https://arxiv.org/abs/hep-ph/9703417}{{\ttfamily hep-ph/9703417}}].

\bibitem{Altarelli:2001ji}
G.~Altarelli, R.~D. Ball and S.~Forte, \emph{{Factorization and resummation of
  small x scaling violations with running coupling}},
  \href{https://doi.org/10.1016/S0550-3213(01)00563-6}{\emph{Nucl. Phys. B}
  {\bfseries 621} (2002) 359}
  [\href{https://arxiv.org/abs/hep-ph/0109178}{{\ttfamily hep-ph/0109178}}].

\bibitem{Altarelli:2003hk}
G.~Altarelli, R.~D. Ball and S.~Forte, \emph{{An Anomalous dimension for small
  x evolution}},
  \href{https://doi.org/10.1016/j.nuclphysb.2003.09.040}{\emph{Nucl. Phys. B}
  {\bfseries 674} (2003) 459}
  [\href{https://arxiv.org/abs/hep-ph/0306156}{{\ttfamily hep-ph/0306156}}].

\bibitem{Altarelli:2005ni}
G.~Altarelli, R.~D. Ball and S.~Forte, \emph{{Perturbatively stable resummed
  small x evolution kernels}},
  \href{https://doi.org/10.1016/j.nuclphysb.2006.01.046}{\emph{Nucl. Phys. B}
  {\bfseries 742} (2006) 1}
  [\href{https://arxiv.org/abs/hep-ph/0512237}{{\ttfamily hep-ph/0512237}}].

\bibitem{Altarelli:2008aj}
G.~Altarelli, R.~D. Ball and S.~Forte, \emph{{Small x Resummation with Quarks:
  Deep-Inelastic Scattering}},
  \href{https://doi.org/10.1016/j.nuclphysb.2008.03.003}{\emph{Nucl. Phys. B}
  {\bfseries 799} (2008) 199}
  [\href{https://arxiv.org/abs/0802.0032}{{\ttfamily 0802.0032}}].

\bibitem{Thorne:1999sg}
R.~S. Thorne, \emph{{Explicit calculation of the running coupling BFKL
  anomalous dimension}},
  \href{https://doi.org/10.1016/S0370-2693(00)00019-8}{\emph{Phys. Lett. B}
  {\bfseries 474} (2000) 372}
  [\href{https://arxiv.org/abs/hep-ph/9912284}{{\ttfamily hep-ph/9912284}}].

\bibitem{Thorne:1999rb}
R.~S. Thorne, \emph{{NLO BFKL equation, running coupling and renormalization
  scales}}, \href{https://doi.org/10.1103/PhysRevD.60.054031}{\emph{Phys. Rev.
  D} {\bfseries 60} (1999) 054031}
  [\href{https://arxiv.org/abs/hep-ph/9901331}{{\ttfamily hep-ph/9901331}}].

\bibitem{Thorne:2001nr}
R.~S. Thorne, \emph{{The Running coupling BFKL anomalous dimensions and
  splitting functions}},
  \href{https://doi.org/10.1103/PhysRevD.64.074005}{\emph{Phys. Rev. D}
  {\bfseries 64} (2001) 074005}
  [\href{https://arxiv.org/abs/hep-ph/0103210}{{\ttfamily hep-ph/0103210}}].

\bibitem{White:2006yh}
C.~D. White and R.~S. Thorne, \emph{{A Global Fit to Scattering Data with NLL
  BFKL Resummations}},
  \href{https://doi.org/10.1103/PhysRevD.75.034005}{\emph{Phys. Rev. D}
  {\bfseries 75} (2007) 034005}
  [\href{https://arxiv.org/abs/hep-ph/0611204}{{\ttfamily hep-ph/0611204}}].

\bibitem{Rothstein:2016bsq}
I.~Z. Rothstein and I.~W. Stewart, \emph{{An Effective Field Theory for Forward
  Scattering and Factorization Violation}},
  \href{https://doi.org/10.1007/JHEP08(2016)025}{\emph{JHEP} {\bfseries 08}
  (2016) 025} [\href{https://arxiv.org/abs/1601.04695}{{\ttfamily
  1601.04695}}].

\bibitem{Catani:1990xk}
S.~Catani, M.~Ciafaloni and F.~Hautmann, \emph{{GLUON CONTRIBUTIONS TO SMALL x
  HEAVY FLAVOR PRODUCTION}},
  \href{https://doi.org/10.1016/0370-2693(90)91601-7}{\emph{Phys. Lett. B}
  {\bfseries 242} (1990) 97}.

\bibitem{Catani:1990eg}
S.~Catani, M.~Ciafaloni and F.~Hautmann, \emph{{High-energy factorization and
  small x heavy flavor production}},
  \href{https://doi.org/10.1016/0550-3213(91)90055-3}{\emph{Nucl. Phys. B}
  {\bfseries 366} (1991) 135}.

\bibitem{Catani:1994sq}
S.~Catani and F.~Hautmann, \emph{{High-energy factorization and small x deep
  inelastic scattering beyond leading order}},
  \href{https://doi.org/10.1016/0550-3213(94)90636-X}{\emph{Nucl. Phys. B}
  {\bfseries 427} (1994) 475}
  [\href{https://arxiv.org/abs/hep-ph/9405388}{{\ttfamily hep-ph/9405388}}].

\bibitem{Bonvini:2019wxf}
M.~Bonvini and F.~Giuli, \emph{{A new simple PDF parametrization: improved
  description of the HERA data}},
  \href{https://doi.org/10.1140/epjp/i2019-12872-x}{\emph{Eur. Phys. J. Plus}
  {\bfseries 134} (2019) 531}
  [\href{https://arxiv.org/abs/1902.11125}{{\ttfamily 1902.11125}}].

\bibitem{Bonvini:2016wki}
M.~Bonvini, S.~Marzani and T.~Peraro, \emph{{Small-$x$ resummation from HELL}},
  \href{https://doi.org/10.1140/epjc/s10052-016-4445-6}{\emph{Eur. Phys. J. C}
  {\bfseries 76} (2016) 597}
  [\href{https://arxiv.org/abs/1607.02153}{{\ttfamily 1607.02153}}].

\bibitem{Bonvini:2017ogt}
M.~Bonvini, S.~Marzani and C.~Muselli, \emph{{Towards parton distribution
  functions with small-$x$ resummation: HELL 2.0}},
  \href{https://doi.org/10.1007/JHEP12(2017)117}{\emph{JHEP} {\bfseries 12}
  (2017) 117} [\href{https://arxiv.org/abs/1708.07510}{{\ttfamily
  1708.07510}}].

\bibitem{Bonvini:2018xvt}
M.~Bonvini and S.~Marzani, \emph{{Four-loop splitting functions at small $x$}},
  \href{https://doi.org/10.1007/JHEP06(2018)145}{\emph{JHEP} {\bfseries 06}
  (2018) 145} [\href{https://arxiv.org/abs/1805.06460}{{\ttfamily
  1805.06460}}].

\bibitem{Bonvini:2018iwt}
M.~Bonvini, \emph{{Small-$x$ phenomenology at the LHC and beyond: HELL 3.0 and
  the case of the Higgs cross section}},
  \href{https://doi.org/10.1140/epjc/s10052-018-6315-x}{\emph{Eur. Phys. J. C}
  {\bfseries 78} (2018) 834}
  [\href{https://arxiv.org/abs/1805.08785}{{\ttfamily 1805.08785}}].

\bibitem{McGowan:2022nag}
J.~McGowan, T.~Cridge, L.~A. Harland-Lang and R.~S. Thorne, \emph{{Approximate
  N$^{3}$LO parton distribution functions with theoretical uncertainties:
  MSHT20aN$^3$LO PDFs}},
  \href{https://doi.org/10.1140/epjc/s10052-023-11236-0}{\emph{Eur. Phys. J. C}
  {\bfseries 83} (2023) 185}
  [\href{https://arxiv.org/abs/2207.04739}{{\ttfamily 2207.04739}}].

\bibitem{BCDMS:1989qop}
{\scshape BCDMS} collaboration, A.~C. Benvenuti et~al., \emph{{A High
  Statistics Measurement of the Proton Structure Functions F(2) (x, Q**2) and R
  from Deep Inelastic Muon Scattering at High Q**2}},
  \href{https://doi.org/10.1016/0370-2693(89)91637-7}{\emph{Phys. Lett. B}
  {\bfseries 223} (1989) 485}.

\bibitem{NewMuon:1996uwk}
{\scshape New Muon} collaboration, M.~Arneodo et~al., \emph{{Accurate
  measurement of F2(d) / F2(p) and R**d - R**p}},
  \href{https://doi.org/10.1016/S0550-3213(96)00673-6}{\emph{Nucl. Phys. B}
  {\bfseries 487} (1997) 3}
  [\href{https://arxiv.org/abs/hep-ex/9611022}{{\ttfamily hep-ex/9611022}}].

\bibitem{CHORUS:2005cpn}
{\scshape CHORUS} collaboration, G.~Onengut et~al., \emph{{Measurement of
  nucleon structure functions in neutrino scattering}},
  \href{https://doi.org/10.1016/j.physletb.2005.10.062}{\emph{Phys. Lett. B}
  {\bfseries 632} (2006) 65}.

\bibitem{NewMuon:1991hlj}
{\scshape New Muon} collaboration, P.~Amaudruz et~al., \emph{{The Gottfried sum
  from the ratio F2(n) / F2(p)}},
  \href{https://doi.org/10.1103/PhysRevLett.66.2712}{\emph{Phys. Rev. Lett.}
  {\bfseries 66} (1991) 2712}.

\bibitem{Gottfried:1967kk}
K.~Gottfried, \emph{{Sum rule for high-energy electron - proton scattering}},
  \href{https://doi.org/10.1103/PhysRevLett.18.1174}{\emph{Phys. Rev. Lett.}
  {\bfseries 18} (1967) 1174}.

\bibitem{Niegawa:1974hk}
A.~Niegawa and K.~Sasaki, \emph{{Adler Sum Rule and Quark Parton Distribution
  Functions in Nucleon}}, \href{https://doi.org/10.1143/PTP.54.192}{\emph{Prog.
  Theor. Phys.} {\bfseries 54} (1975) 192}.

\bibitem{Field:1976ve}
R.~D. Field and R.~P. Feynman, \emph{{Quark Elastic Scattering as a Source of
  High Transverse Momentum Mesons}},
  \href{https://doi.org/10.1103/PhysRevD.15.2590}{\emph{Phys. Rev. D}
  {\bfseries 15} (1977) 2590}.

\bibitem{SeaQuest:2021zxb}
{\scshape SeaQuest} collaboration, J.~Dove et~al., \emph{{The asymmetry of
  antimatter in the proton}},
  \href{https://doi.org/10.1038/s41586-022-04707-z}{\emph{Nature} {\bfseries
  590} (2021) 561} [\href{https://arxiv.org/abs/2103.04024}{{\ttfamily
  2103.04024}}].

\bibitem{FNALE906:2022xdu}
{\scshape FNAL E906, SeaQuest} collaboration, J.~Dove et~al.,
  \emph{{Measurement of flavor asymmetry of the light-quark sea in the proton
  with Drell-Yan dimuon production in p+p and p+d collisions at 120 GeV}},
  \href{https://doi.org/10.1103/PhysRevC.108.035202}{\emph{Phys. Rev. C}
  {\bfseries 108} (2023) 035202}
  [\href{https://arxiv.org/abs/2212.12160}{{\ttfamily 2212.12160}}].

\bibitem{NuSea:2001idv}
{\scshape NuSea} collaboration, R.~S. Towell et~al., \emph{{Improved
  measurement of the anti-d / anti-u asymmetry in the nucleon sea}},
  \href{https://doi.org/10.1103/PhysRevD.64.052002}{\emph{Phys. Rev. D}
  {\bfseries 64} (2001) 052002}
  [\href{https://arxiv.org/abs/hep-ex/0103030}{{\ttfamily hep-ex/0103030}}].

\bibitem{CDF:2010vek}
{\scshape CDF} collaboration, T.~A. Aaltonen et~al., \emph{{Measurement of
  $d\sigma/dy$ of Drell-Yan $e^+e^-$ pairs in the $Z$ Mass Region from
  $p\bar{p}$ Collisions at $\sqrt{s}=1.96$ TeV}},
  \href{https://doi.org/10.1016/j.physletb.2010.06.043}{\emph{Phys. Lett. B}
  {\bfseries 692} (2010) 232}
  [\href{https://arxiv.org/abs/0908.3914}{{\ttfamily 0908.3914}}].

\bibitem{D0:2007djv}
{\scshape D0} collaboration, V.~M. Abazov et~al., \emph{{Measurement of the
  Shape of the Boson Rapidity Distribution for $p \bar{p} \to Z / \gamma^* \to
  e^{+} e^{-} + X$ Events Produced at $\sqrt{s}$ of 1.96-TeV}},
  \href{https://doi.org/10.1103/PhysRevD.76.012003}{\emph{Phys. Rev. D}
  {\bfseries 76} (2007) 012003}
  [\href{https://arxiv.org/abs/hep-ex/0702025}{{\ttfamily hep-ex/0702025}}].

\bibitem{Ball:2013lla}
{\scshape NNPDF} collaboration, R.~D. Ball, S.~Forte, A.~Guffanti, E.~R.
  Nocera, G.~Ridolfi and J.~Rojo, \emph{{Unbiased determination of polarized
  parton distributions and their uncertainties}},
  \href{https://doi.org/10.1016/j.nuclphysb.2013.05.007}{\emph{Nucl. Phys. B}
  {\bfseries 874} (2013) 36} [\href{https://arxiv.org/abs/1303.7236}{{\ttfamily
  1303.7236}}].

\bibitem{Nocera:2014gqa}
{\scshape NNPDF} collaboration, E.~R. Nocera, R.~D. Ball, S.~Forte, G.~Ridolfi
  and J.~Rojo, \emph{{A first unbiased global determination of polarized PDFs
  and their uncertainties}},
  \href{https://doi.org/10.1016/j.nuclphysb.2014.08.008}{\emph{Nucl. Phys. B}
  {\bfseries 887} (2014) 276}
  [\href{https://arxiv.org/abs/1406.5539}{{\ttfamily 1406.5539}}].

\bibitem{deFlorian:2009vb}
D.~de~Florian, R.~Sassot, M.~Stratmann and W.~Vogelsang, \emph{{Extraction of
  Spin-Dependent Parton Densities and Their Uncertainties}},
  \href{https://doi.org/10.1103/PhysRevD.80.034030}{\emph{Phys. Rev. D}
  {\bfseries 80} (2009) 034030}
  [\href{https://arxiv.org/abs/0904.3821}{{\ttfamily 0904.3821}}].

\bibitem{Ethier:2017zbq}
J.~J. Ethier, N.~Sato and W.~Melnitchouk, \emph{{First simultaneous extraction
  of spin-dependent parton distributions and fragmentation functions from a
  global QCD analysis}},
  \href{https://doi.org/10.1103/PhysRevLett.119.132001}{\emph{Phys. Rev. Lett.}
  {\bfseries 119} (2017) 132001}
  [\href{https://arxiv.org/abs/1705.05889}{{\ttfamily 1705.05889}}].

\bibitem{DeFlorian:2019xxt}
D.~De~Florian, G.~A. Lucero, R.~Sassot, M.~Stratmann and W.~Vogelsang,
  \emph{{Monte Carlo sampling variant of the DSSV14 set of helicity parton
  densities}}, \href{https://doi.org/10.1103/PhysRevD.100.114027}{\emph{Phys.
  Rev. D} {\bfseries 100} (2019) 114027}
  [\href{https://arxiv.org/abs/1902.10548}{{\ttfamily 1902.10548}}].

\bibitem{Zhou:2022wzm}
{\scshape Jefferson Lab Angular Momentum (JAM)} collaboration, Y.~Zhou, N.~Sato
  and W.~Melnitchouk, \emph{{How well do we know the gluon polarization in the
  proton?}}, \href{https://doi.org/10.1103/PhysRevD.105.074022}{\emph{Phys.
  Rev. D} {\bfseries 105} (2022) 074022}
  [\href{https://arxiv.org/abs/2201.02075}{{\ttfamily 2201.02075}}].

\bibitem{Cocuzza:2022jye}
{\scshape Jefferson Lab Angular Momentum (JAM)} collaboration, C.~Cocuzza,
  W.~Melnitchouk, A.~Metz and N.~Sato, \emph{{Polarized antimatter in the
  proton from a global QCD analysis}},
  \href{https://doi.org/10.1103/PhysRevD.106.L031502}{\emph{Phys. Rev. D}
  {\bfseries 106} (2022) L031502}
  [\href{https://arxiv.org/abs/2202.03372}{{\ttfamily 2202.03372}}].

\bibitem{Karpie:2023nyg}
J.~Karpie, R.~M. Whitehill, W.~Melnitchouk, C.~Monahan, K.~Orginos, J.~W. Qiu
  et~al., \emph{{Gluon helicity from global analysis of experimental data and
  lattice QCD Ioffe time distributions}},
  [\href{https://arxiv.org/abs/2310.18179}{{\ttfamily 2310.18179}}].

\bibitem{Hekhorn:2024tqm}
F.~Hekhorn, G.~Magni, E.~R. Nocera, T.~R. Rabemananjara, J.~Rojo, A.~Schaus
  et~al., \emph{{Heavy Quarks in Polarised Deep-Inelastic Scattering at the
  Electron-Ion Collider}},  [\href{https://arxiv.org/abs/2401.10127}{{\ttfamily
  2401.10127}}].

\bibitem{deFlorian:2024utd}
D.~de~Florian, S.~Forte and W.~Vogelsang, \emph{{Higgs production at RHIC and
  the positivity of the gluon helicity distribution}},
  [\href{https://arxiv.org/abs/2401.10814}{{\ttfamily 2401.10814}}].

\bibitem{ParticleDataGroup:2022pth}
{\scshape Particle Data Group} collaboration, R.~L. Workman et~al.,
  \emph{{Review of Particle Physics}},
  \href{https://doi.org/10.1093/ptep/ptac097}{\emph{PTEP} {\bfseries 2022}
  (2022) 083C01}.

\bibitem{Bertone:2017gds}
V.~Bertone, \emph{{APFEL++: A new PDF evolution library in C++}},
  \href{https://doi.org/10.22323/1.297.0201}{\emph{PoS} {\bfseries DIS2017}
  (2018) 201} [\href{https://arxiv.org/abs/1708.00911}{{\ttfamily
  1708.00911}}].

\bibitem{Bourrely:2007if}
C.~Bourrely, J.~Soffer and F.~Buccella, \emph{{Strangeness asymmetry of the
  nucleon in the statistical parton model}},
  \href{https://doi.org/10.1016/j.physletb.2007.02.063}{\emph{Phys. Lett. B}
  {\bfseries 648} (2007) 39}
  [\href{https://arxiv.org/abs/hep-ph/0702221}{{\ttfamily hep-ph/0702221}}].

\end{thebibliography}\endgroup

\end{document}